\documentclass[
 reprint,
 nofootinbib,
 amsmath, amssymb,
 aps,
]{revtex4-2}

\usepackage{graphicx}
\usepackage{xcolor}
\usepackage{dcolumn}
\usepackage{bm}
\usepackage{comment}
\usepackage{hyperref}

\begin{document}

\preprint{APS/123-QED}

\title{Consensus in Effective Model Inference for Disordered Ising Systems}

\author{Ashveed Ashraf}
\affiliation{Department of Physics,
\href{https://ror.org/0440p1d37}{Gandhi Institute of Technology and Management (GITAM) University}, Bengaluru, India}

\date{\today}

\begin{abstract}

When independent learners are trained on data from the same disordered system, finite and noisy samples can lead them to different descriptions of the underlying physics. We quantify the reproducibility of such descriptions through consensus, the degree of agreement among independently trained models. In a teacher–student framework on the two-dimensional random-bond Ising model, an ensemble of students each infers a single uniform effective coupling from a finite set of equilibrium configurations drawn from one quenched bond realization. Aggregating the inferred couplings across students and realizations yields the consensus distribution, whose width quantifies the agreement among learners. The variance of this distribution separates exactly into a finite-sampling contribution, bounded below by inverse the of Fisher information, and a quenched-disorder contribution. Sum of these two variance is minimized near the critical region, so the consensus width has a minimum near the pseudocritical temperature that sharpens with increasing disorder. The two minima arise from different mechanisms: the sampling term inherits the critical peak of the Fisher information, while the disorder term is temperature-independent at leading order and acquires its critical minimum only at fourth order in the disorder strength, through the covariance between the linear and cubic responses to the bond disorder. Because the students fit a uniform coupling to a heterogeneous lattice, the inferred coupling also carries a misspecification bias that survives in the infinite-data limit. This bias displaces the minimum of the mean-squared error above the consensus minimum, so the temperature at which independent learners agree most closely is not the temperature at which their shared answer is most faithful.

\end{abstract}

\maketitle

\section{\label{sec:level1}INTRODUCTION}

Reconstructing interactions in a complex system from observations
of its states is a problem shared by statistical physics, machine
learning, and quantitative biology~\cite{Chayes1984,Nguyen_2017,Kunkin1969,Changlani2015,Krumsiek2011}. As experiments across scales become increasingly accessible, inverse problems have drawn growing interest. A canonical instance is the inverse Ising problem, where the couplings $J_{ij}$ and fields $h_i$ are reconstructed from the spin configurations $\{\sigma_i\}$ that generated them. The same formal problem underlies the reconstruction of neural connectivity~\cite{Cocco2009,Roudi2009}, protein structure determination \cite{Wang2017}, inference of gene-regulatory interactions\cite{Lezon2006}, and modeling of collective behavior in
biological populations\cite{Bialek}. There exists different tool kits for the inference problems differing in cost and validation regimes \cite{Kappen,NguyenBerg2012,Nguyen_Berg_2012,Cocco_2011,Hyvrinen2006ConsistencyOP} but share a common objective to infer the parameters of the generating model.

The success of such inference is governed by \emph{accuracy} by comparing inferred parameters against the true ones. In order for that, a well-defined ground truth should exist, and the fitted model must be rich enough to represent the ground truth. In practice, neither condition holds. Real interactions are heterogeneous, yet the models fitted to them are
deliberately simple. This is because the true distribution is usually unknown, moreover, a single effective parameter is easier to interpret than a full set of parameters, and sometimes limited data cannot constrain a richer model. For such a quantity, accuracy alone is an incomplete criterion as there exists no unique ground truth. So one must also ask whether independent analyses agree on the inferred value or more clearly whether the effective description is reproducible.

We make this notion precise through the idea of \emph{consensus}.
Consider many independent learners, each given data from the
same class of disordered system and each fitting the same coarse
effective model. Due to different data from different disordered realizations of the same effective model, the inferred effective parameters form a distribution rather than a single value. We call this the
\emph{consensus distribution}, and width of the distribution as \emph{consensus width}. A narrow distribution means the effective description is robust, whereas a broad one means the inferred parameter is largely an artifact of the particular dataset or the disorder realization. The same construction, training an ensemble of independent models and reading the spread of their predictions, is used in machine learning as a practical measure of predictive uncertainty~\cite{Lakshminarayanan2017,Fort2019}. Consensus, however, is distinct from accuracy as a misspecified learner may yield highly consistent estimates across datasets while deviating from the true parameters. Thus, the conditions that promote reproducibility need not coincide with those that minimize inference error. A central goal of this work is therefore to characterize the conditions under which reproducibility and accuracy emerge, and to understand the trade-off between them.

For an inverse Ising problem, temperature is the natural control for answering this question. At low temperature, system carries lesser information about the couplings while at high temperature, thermal noise washes out the correlations that encode them. Earlier work has shown that inference accuracy maximizes in an intermediate, just below the critical regime, where the configurations are both varied and strongly correlated \cite{Ngampruetikorn_2022}.

Previous studies, have been done on clean Ising model with uniform couplings, or introduce disorder in the system by removing the fewer bond couplings as in random graphs. But quenched disorder adds a further complication to the inference problem. We therefore study the random-bond Ising model, with each bond drawn independently from a continuous distribution and this setting produces a genuine spread of inferred effective couplings across realizations.

We address this with a teacher-student model ~\cite{Loureiro_2022,Zdeborov__2016} for the
two-dimensional random-bond Ising model. A \emph{teacher} is one
quenched bond realization $\omega$ held at temperature $T$; from it we
draw equilibrium configurations and distribute them among independent
\emph{students}. Each student sees only finite number of configurations and, rather than attempting to recover the full disordered couplings, fits a
single uniform effective coupling. We denote the inferred coupling by the student $s$ with finite number of data for the realization $\omega$ as $\hat{J}_{\omega,s}$ whereas $J_\omega^*$ is the infinite data effective coupling of realization $\omega$. 
\begin{equation}
    \hat{J}_{\omega,s} \xrightarrow{M \to \infty} J^*_{\omega}
    \label{eq:1}
\end{equation}
Here, data represents equilibrium configurations of the disordered Ising model. The uniform coupling each student returns is a compressed effective description of the disordered lattice, collected across many students and many realizations, these couplings form the consensus distribution $P(\hat{J}_{\omega,s})$ for this disordered model.

The central structural result is that the variance of this distribution
separates cleanly into two physically distinct contributions given by the law of total variance as 
\begin{equation}
V_{\rm consensus}
=
\mathbb E_\omega\!\big[\operatorname{Var}_s(\hat J_{\omega,s}\mid\omega)\big]
+
\operatorname{Var}_\omega\big[\mathbb E_{s}(\hat{J}_{\omega,s}\mid\omega)\big]
\label{eq:2}
\end{equation}
The first term on the RHS is sampling noise, the disagreement among students who share a teacher but see different finite samples, denoted by $V_{\mathrm{samples}}$. The second term is quenched-disorder variance, the genuine spread of the infinite-data effective coupling across realizations, denoted by $V_{\mathrm{disorder}}$. The sum of the two gives the $V_{\mathrm{consensus}}$, variance of the distribution $P(\hat{J}_{\omega,s})$ across all students and teachers. Reproducibility of the inferred coupling is controlled by their sum, and each have entirely different dependencies on temperature. Analyzing them separately is one of the main aims of this work.

We first analyze the sampling variance, find a lower bound for the sampling error and locate the temperature at which it is minimized. We then turn to the disorder variance, identify its own minimum, and reason for the emergence of the minimum. Combining the two gives the consensus variance, whose behavior across temperature and disorder strength provides the regime in which independent learners agree most closely. Finally, we bring accuracy and reproducibility together through the mean-squared error, which adds to the consensus variance a bias term measuring the deviation of the inferred coupling from the mean bond strength. This bias vanishes for the clean Ising model, and varies with temperature and grows stronger with disorder. We find that the consensus width is narrowest in the critical region and sharpens as the disorder grows. The mean-squared error optimum, however, is displaced above the critical region by the bias, so the temperature at which independent learners agree most closely is not the temperature at which their shared answer is most faithful.

\section{Model and data}
\label{sec:model}

We consider the two-dimensional random-bond Ising model on an $L \times L$
square lattice with $L = 50$ with periodic boundary conditions. The
Hamiltonian is
\begin{equation}
    \mathcal{H} = -\sum_{\langle i,j\rangle} J_{ij}\,\sigma_i\sigma_j,
    \qquad \sigma_i = \pm 1,
    \label{eq:3}
\end{equation}
with the sum over nearest-neighbour pairs. For each disorder strength
$\delta \in [0, 1)$ the couplings are drawn independently from a
uniform distribution,
\begin{equation}
    J_{ij} \sim \mathcal{U}(J_0 - \delta,\, J_0 + \delta),
    \qquad J_0 = 1,
    \label{eq:4}
\end{equation}
so that all bonds remain ferromagnetic and no frustration arises. Each
realization of the bond variables $\{J_{ij}^\omega\}$ defines a
distinct teacher labelled by $\omega$; for each $\delta$ we generate
$N_\omega = 200$ independent realizations.

For each pair $(\omega, T)$, equilibrium spin configurations are
sampled from the Boltzmann distribution denoting the teacher as
\begin{equation}
    P_{\mathrm{\omega}}\!\left(\{\sigma_i\}\right)
    = \frac{1}{Z(\{J_{ij}^{\omega}\})}
    \exp\!\left(\beta\sum_{\langle ij \rangle}
    J_{ij}^{\omega}\,\sigma_i\sigma_j\right),
    \label{eq:5}
\end{equation}
with $\beta = 1/T$, $\{J_{ij}\}^{\omega}$ are the quenched couplings of realization $\omega$, and $Z(\{J_{ij}^{\omega}\}) = \sum_{\{\sigma_i\}}
\exp\!\bigl(\beta\sum_{\langle ij\rangle} J_{ij}^{\omega}\sigma_i\sigma_j\bigr)$ is the partition function. We span the range $T \in [1.60, 3.00]$ in steps of $\Delta T = 0.04$, covering the ordered phase, the critical region, and the disordered phase. At each $(\omega, T)$ we collect $N_{\mathrm{tot}} = 25000$ configurations, stored at intervals of $10\tau_{\rm int}$, where $\tau_{\rm int}$ is the measured integrated autocorrelation time. The complete Monte Carlo scheme is given in
Appendix~\ref{app:MC}.

Rather than storing the full $L^2$ spin configurations, we encode each
configuration through its local pattern statistics. On a square
lattice, every spin has four nearest neighbors, yielding $2^4 = 16$
distinct neighbourhood patterns indexed by $x = 0, 1, \ldots, 15$.
For each configuration we record the number of sites $n_+(x)$ and
$n_-(x)$ at which pattern $x$ occurs with central spin $+1$ and $-1$ respectively, compressing the configuration into a 32-component
vector. Because the conditional distribution of a spin given the rest of the lattice depends only on its four neighbours, these counts are sufficient statistics for the pseudolikelihood~\cite{Besag1974}: they determine it exactly, while reducing the configuration dimensionality from $L^2 = 2500$ spins to $32$ counts. See Appendix~\ref{sec:patterns} for details.

\section{Consensus learning framework}
\label{sec:consensus_framework}

We now formalize the notion of consensus within a teacher-student
framework. For each disorder realization $\omega$ and temperature $T$,
the teacher represented by Eq.~\eqref{eq:5}, generates a dataset of
$N_\mathrm{tot}$ equilibrium configurations for a given $\omega$,
\begin{equation}
    \mathcal{D}_\omega = \bigl\{\{\sigma_i\}^{(1)}, \{\sigma_i\}^{(2)}, \ldots,
    \{\sigma_i\}^{(N_\mathrm{tot})}\bigr\}
    \sim P_{\omega}\!\left(\{\sigma_i\}\mid\{J_{ij}^{\omega}\}\right),
    \label{eq:6}
\end{equation}
which is partitioned among $N_s$ students, each receiving a finite
subset $\mathcal{D}_{\omega,s} \subset \mathcal{D}_\omega$ of $M$
configurations. From its subset, each student infer a single uniform effective coupling. Aggregating these inferred couplings across students and disorder realizations yields the consensus distribution. The following subsections describe the hypothesis class, which is defined as the set of functions within which each student performs the inference optimization and the inference procedure step by step along with the quantities that we use to characterize the consensus distribution. 

\subsection{Student inference}
\label{sec:student_pipeline}

The inference problem considered here is different from the
reconstruction of all bonds in the Ising lattice. A disordered teacher contains $2L^2$ bond parameters, whereas each student is restricted to a
one-parameter clean Ising model with uniform coupling.

For a central site $i$, we denote its four-neighbor spin pattern by
\begin{equation}
    x = (\sigma_1,\sigma_2,\sigma_3,\sigma_4),
    \label{eq:7}
\end{equation}
There are sixteen such possible neighborhood patterns. The sum of the neighboring spins is given by
\begin{equation}
    S(x) = \sum_{k=1}^{4} \sigma_k
    \in \{-4,-2,0,2,4\}.
    \label{eq:8}
\end{equation}
The conditional distribution of a single spin depends on the rest of the lattice only through its four neighbors, and takes the logistic form given by
\begin{equation}
    P_\omega\big(\sigma_i = +1 \mid \{\sigma_j\}\big) = \frac{1}{1+e^{-2h_i}},
    \qquad
    h_i = \beta\sum_{j\in\partial i} J_{ij}\,\sigma_j .
    \label{eq:9}
\end{equation}
We call $h_i$ the \emph{local field} at site $i$. If all bonds are equal to a common value $J$, then $h_i$ depends on the neighborhood only through $S(x)$, and $f(x) = \beta J S(x)$. We define student’s hypothesis class as the corresponding one-parameter family of fields,
\begin{equation}
    \mathcal{F}_{\mathrm{student}}
    = \bigl\{\, f_\omega(x) = \beta J\, S(x) \;:\; J \in \mathbb{R} \,\bigr\} .
    \label{eq:10}
\end{equation}
 At $\delta = 0$, teacher is a clean Ising model and belongs to this family. Then in the limit of infinite data, the inferred effective coupling value will be $J_0$, the true parameter. At finite disorder, however, the exact local field depends on the individual surrounding bonds and cannot in general be represented by a single uniform coupling. The student's task is therefore a compression of heterogeneous model onto the one-dimensional model $\mathcal{F_{\mathrm{student}}}$.

The cost of this compression is the gap between the loss the student attains within its restricted one-parameter family and the loss attained over the full class of pattern functions, which represents the teacher exactly. Let
\begin{equation}
    \mathcal{L}_\omega(J) = \sum_x p_\omega(x)
    \big[f_\omega(x) - \beta J S(x)\big]^2
    \label{eq:11}
\end{equation}
be the population loss of a candidate coupling $J$ for teacher $\omega$. Here, population loss means the loss under the true distribution, computed from a student's $M$ configurations, such that $M \to \infty$. We define $p_\omega(x)$ as probability that a site has neighborhood pattern $x$, such that $\sum_x p_\omega(x) = 1$. Write
$\mathcal{L}^*_{\mathcal{F}} = \min_J \mathcal{L}_\omega(J) =
\mathcal{L}_\omega(J^*_\omega)$ for the best loss attainable within the student's hypothesis class $\mathcal{F}_{\rm student}$, and $\mathcal{L}^*_{\rm full}$ for the best attainable loss over the unrestricted class of functions. The excess loss of the coupling the student actually infers from its $M$ configurations then admits a decomposition into two physically distinct
contributions,
\begin{equation}
    \underbrace{\mathcal{L}_\omega(\hat J_{\omega,s})
    - \mathcal{L}^*_{\rm full}}_{\text{excess loss}}
    =
    \underbrace{\big(\mathcal{L}^*_{\mathcal{F}}
    - \mathcal{L}^*_{\rm full}\big)}_{\text{approximation error}}
    +
    \underbrace{\big(\mathcal{L}_\omega(\hat J_{\omega,s})
    - \mathcal{L}^*_{\mathcal{F}}\big)}_{\text{estimation error}},
    \label{eq:12}
\end{equation}
both non-negative: the first because
$\mathcal{F}_{\rm student}\subset\mathcal{F}_{\rm full}$, the second because
$J^*_\omega$ minimizes $\mathcal{L}_\omega$~\cite{White1982}.

The estimation error arises from finite sampling. With sufficient data the
student converges to the best uniform fit, $\hat J_{\omega,s} \to J^*_\omega$, so this contribution vanishes as $M\to\infty$ regardless of disorder strength. The approximation error arises from the mismatch between the student's hypothesis class and the teacher. At $\delta = 0$ the teacher is uniform and lies inside $\mathcal{F}_{\rm student}$, so it vanishes; but at $\delta > 0$ no single coupling can reproduce heterogeneous bonds and it remains strictly positive. This contribution cannot be removed by collecting more samples and is reduced only by enlarging the hypothesis class which can represent the disordered teacher. This mismatch is the origin of the bias analyzed in Sec.~\ref{sec:results}.

To realize this in practice, the teacher first generates $N_\mathrm{tot} = 25000$ equilibrium configurations for a fixed realization $\omega$ and temperature $T$, which are distributed among $N_s$ students, each receiving $M$ configurations with $N_\mathrm{tot}=N_s\times M$. Each student then infers its effective coupling using the $M$ samples in two steps -  $i$) estimating the local field $f(x)$ from the pattern counts $ii$) projection of the inferred field $f(x)$ onto the uniform Ising model $\mathcal{F_{\mathrm{student}}}$ to yield a single effective coupling $\hat{J}$, as detailed in Sec.~\ref{sec:local_field}.

\begin{figure}[t!]
    \includegraphics[width=\columnwidth]{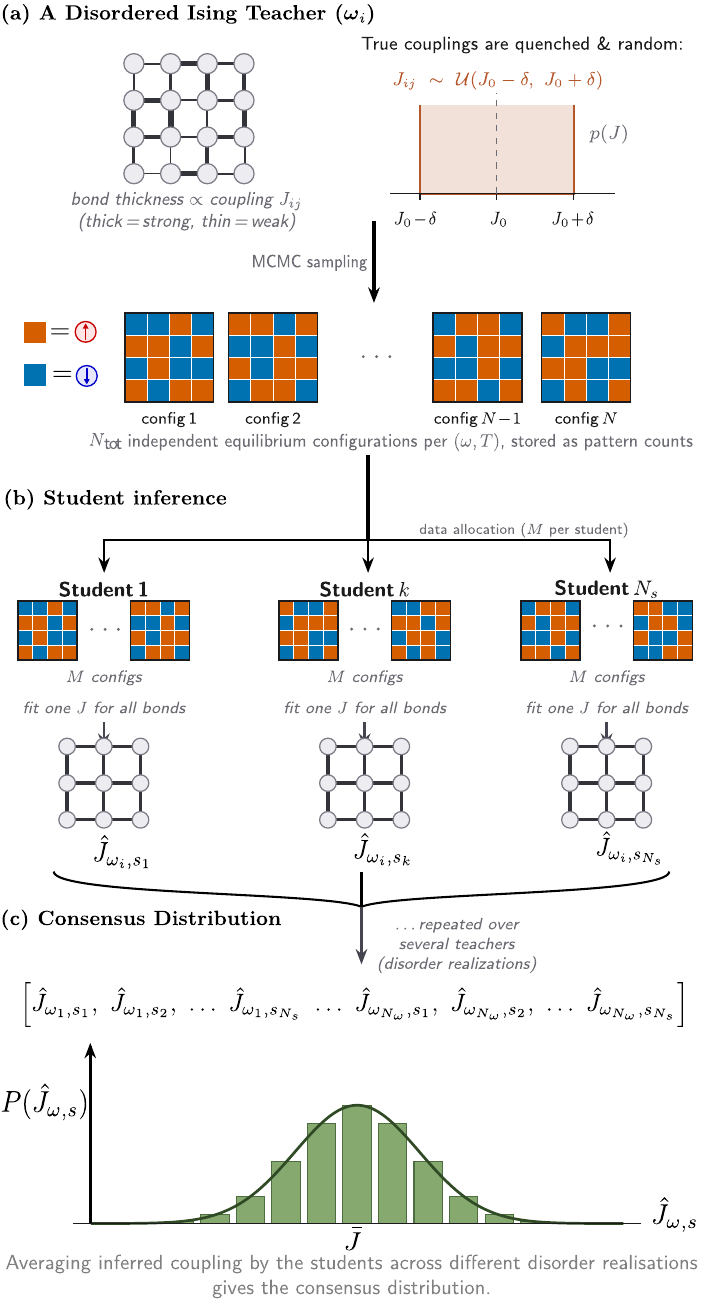}
    \caption{Consensus learning framework.
    \textbf{(a)} A disordered Ising teacher: bonds
    $J_{ij} \sim \mathcal{U}(J_0 - \delta, J_0 + \delta)$ are quenched and
    random for each realization $\omega$. MCMC sampling produces
    $N_\mathrm{tot}$ equilibrium spin configurations per $(\omega, T)$, stored
    as pattern counts.
    \textbf{(b)} Student inference: each of $N_s$ students, denoted by $s$,
    receives $M$ configurations sampled without replacement from the same
    teacher $\omega$ and infers a single effective coupling
    $\hat{J}_{\omega,s}$ by fitting a uniform Ising model.
    \textbf{(c)} Consensus distribution: aggregating $\hat{J}_{\omega,s}$
    across all students and all teachers builds the consensus distribution
    $P(\hat{J}_{\omega,s})$, whose width quantifies how tightly independent
    learners agree on the effective description of the disordered system. Its
    mean is denoted $\bar{J}$.}
    \label{fig:1}
\end{figure}

\subsection{Extraction of the local field}
\label{sec:local_field}

For the 2D Ising model the conditional probability of a central spin given its neighbors is Eq.~\eqref{eq:9}. Writing $h_i = f_\omega(x)$ by assuming all bonds are identical and then inverting the Eq.~\eqref{eq:9} gives 
\begin{equation}
f_\omega(x)
=
\frac12\ln
\frac{P_\omega(\sigma_i=+1\mid x)}
     {P_\omega(\sigma_i=-1\mid x)}.
\label{eq:13}
\end{equation}
$f_\omega(x)$ is the local field inferred from the whole distribution by the student for a given teacher $\omega$. Minimizing the loss function given in Eq.~\eqref{eq:11} gives the inferred coupling. But, for a student receiving finite $M$ configurations, the cumulative counts of patterns are obtained by summing over the configurations in its data subset $\mathcal{D}_{\omega, s}$.

From these counts, the conditional probability that the central spin is $+1$ given neighborhood pattern $x$ is estimated with,
\begin{equation}
    \hat{P}_{\omega}(\sigma = +1 \mid x)
    = \frac{n_+(x) + \alpha_s}{N(x) + 2\alpha_s},
    \label{eq:14}
\end{equation}
where $N(x) = n_+(x) + n_-(x)$ is the total number of times pattern
$x$ was observed, $n_+(x)$ and $n_-(x)$ represent the number of times the patterns is observed with central spin as $+1$ and $-1$ respectively. $\alpha_s > 0$ is a smoothing parameter. Throughout this work, the Jeffreys prior $\alpha_s = 0.5$~\cite{Jeffreys1946} has been used for getting better estimates even for rarely observed patterns.

Here, a student receives only a finite dataset $\mathcal{D}_{\omega,s}$. The inferred coupling by the student from the finite dataset it receives is given by
\begin{equation}
    \hat{J}_{\omega,s} = \arg\min_{J} \sum_{x=0}^{15} N(x)\,
    \bigl[f(x) - \beta J\,S(x)\bigr]^2
    \label{eq:15}
\end{equation}
where $N(x)$ is the pattern weights such that patterns observed more frequently contribute more and $f(x)$ is the finite data limit of local field obtained by combining Eq.~\eqref{eq:13} and Eq.~\eqref{eq:14}. The closed form solution yields
\begin{equation}
    \hat{J}_{\omega,s}
    =  \frac{1}{\beta}\frac{\displaystyle\sum_{x=0}^{15} N(x)\,S(x)\,f(x)}
           {\displaystyle\sum_{x=0}^{15} N(x)\,S(x)^2},
    \label{eq:16}
\end{equation}
$\hat{J}_{\omega, s}$ is the coupling that we infer for each disorder $\delta$ and temperature $T$. In the infinite data limit, for $\delta>0$, $\hat{J}_{\omega,s} \xrightarrow{M \to \infty} J^*_{\omega}$ and for $\delta=0$, $\hat{J}_{\omega,s} \xrightarrow{M \to \infty} J_0$ exactly.

Instead of calculating the finite data local field $f(x)$ using the Eq.~\eqref{eq:9}, we use the kernel ridge regression. From Eq.~\eqref{eq:9} and Eq.~\eqref{eq:13}, it is clear that the inference of $f(x)$ does not assume any functional dependence of $f(x)$ on the neighboring spin configurations: each of the sixteen values is estimated independently from the counts. But using kernels gives the flexibility of choosing the functional form of $f(x)$ by assuming the relationship  between the central spins and the nearest neighboring spins. This helps in splitting the linear and non linear components arising from the interaction between the spins in the system. For the results, we have used the identity kernel which gives the exact same $f(x)$ as mentioned in the Eq.~\eqref{eq:13} in the finite data limit. We have also done the analysis with linear, isotropic and RBF kernels, which are explained in detail in Appendix~\ref{app:kernels_RC}. Fig.~\ref{fig:1} provides a schematic overview of the complete consensus learning framework.

\subsection{Consensus distribution and variance decomposition}
\label{sec:variance_decomposition}

For each disorder realization \(\omega\) and student \(s\), the inference pipeline produces \(\hat J_{\omega,s}\). The collection over students and teachers defines the empirical consensus distribution $P(\hat{J}_{\omega,s})$. The variance of this consensus distribution can be written as,
\begin{equation}
\operatorname{Var}_{\omega,s}(\hat J_{\omega,s})
=
\mathbb E_\omega\!\left[
\operatorname{Var}_{s}(\hat J_{\omega,s}\mid\omega)
\right]
+
\operatorname{Var}_\omega\!\left[
\mathbb E_s(\hat J_{\omega,s}\mid\omega)
\right].
\label{eq:17}
\end{equation}
we denote each term as
\begin{align}
V_{\rm consensus}
&\equiv
\operatorname{Var}_{\omega,s}(\hat{J}_{\omega,s}), \\[6pt]
V_{\rm samples}
&\equiv
\mathbb{E}_\omega\!\left[
    \operatorname{Var}_{s}(\hat{J}_{\omega,s}\mid\omega)
\right], \\[6pt]
V_{\rm disorder}
&\equiv
\operatorname{Var}_\omega\!\left[
\mathbb E_s(\hat J_{\omega,s}\mid\omega)
\right]
\end{align}
$V_{\rm sample}$ captures finite data disagreement among students learning the same teacher while $V_{\rm disorder}$ represents the variation in the inferred coupling across different teachers.

For finite \(N_s\), we estimate the conditional mean and variance for teacher \(\omega\) by
\begin{align}
\bar J_\omega
&=
\frac1{N_s}\sum_{s=1}^{N_s}\hat J_{\omega,s},
\label{eq:21}\\
V_\omega
&=
\frac1{N_s-1}\sum_{s=1}^{N_s}
(\hat J_{\omega,s}-\bar J_\omega)^2.
\label{eq:22}
\end{align}
The empirical sampling variance is
\begin{equation}
 V_{\rm samples}
=
\frac1{N_\omega}\sum_{\omega=1}^{N_\omega}V_\omega.
\label{eq:23}
\end{equation}
Between teachers one can write,
\begin{equation}
 V_{\rm disorder}
=
\frac1{N_\omega-1}
\sum_{\omega=1}^{N_\omega}
(\bar J_\omega-\bar J)^2,
\qquad
\bar J=\frac1{N_\omega}\sum_\omega\bar J_\omega,
\label{eq:24}
\end{equation}
The total variance of the consensus distribution, taken over all the students and teachers can be written as 
\begin{equation}
    V_\mathrm{consensus}
    = \frac{1}{N_\omega N_s }\sum_{\omega,s}
    \bigl(\hat{J}_{\omega,s} - \langle\hat{J} \rangle\bigr)^2.
    \label{eq:25}
\end{equation}

Eq.~\eqref{eq:17} is exact for any $N_s$ and $N_\omega$. The estimator
\eqref{eq:24}, however, is biased upward by $V_{\rm sample}/N_s$, since
$\bar J_\omega$ is itself an average over $N_s$ students; at $N_s = 250$ this is a $0.4\%$ correction. Thus in the large \(N_s,N_\omega\) limit it approaches
\begin{equation}
V_{\rm consensus}
\approx
V_{\rm samples}+V_{\rm disorder}.
\label{eq:26}
\end{equation}
\section{Results}
\label{sec:results}

For each disorder strength $\delta$ and temperature $T$ we pool the couplings
inferred by every student from every teacher realization,
$\{\hat{J}_{\omega,s}\}$ with $\omega = 1,\dots,N_\omega$ labelling disorder
realizations and $s = 1,\dots,N_s$ labelling students, and form the \emph{consensus distribution} $P(\hat{J})$.  The mean of the distribution, $\bar{J}$ is displaced from the true coupling $J_0$ due to the
inference bias, and the width measures how strongly independent learners agree with one another. The central result of this section is that these two quantities are optimized in different thermal regimes. Throughout, $T_c^\mathrm{eff}(\delta)$ denotes the pseudocritical temperature of the finite $L \times L$ Ising system at disorder strength $\delta$, extracted from the peak of the specific heat. For the clean system finite lattice estimate $T_c^{\rm eff}(0) = 2.28$ lies slightly above the exact value $T_c = 2/\ln(1+\sqrt{2}) \simeq 2.269$, as expected for finite lattices. Disorder drives it downward, and by $\delta = 0.999$ it has fallen to $T_c^{\rm eff}(0.999) = 2.04$, well below the clean value~\cite{Harris1974}.

\begin{figure*}[t!]
    \centering
    \includegraphics[width=0.99\linewidth]{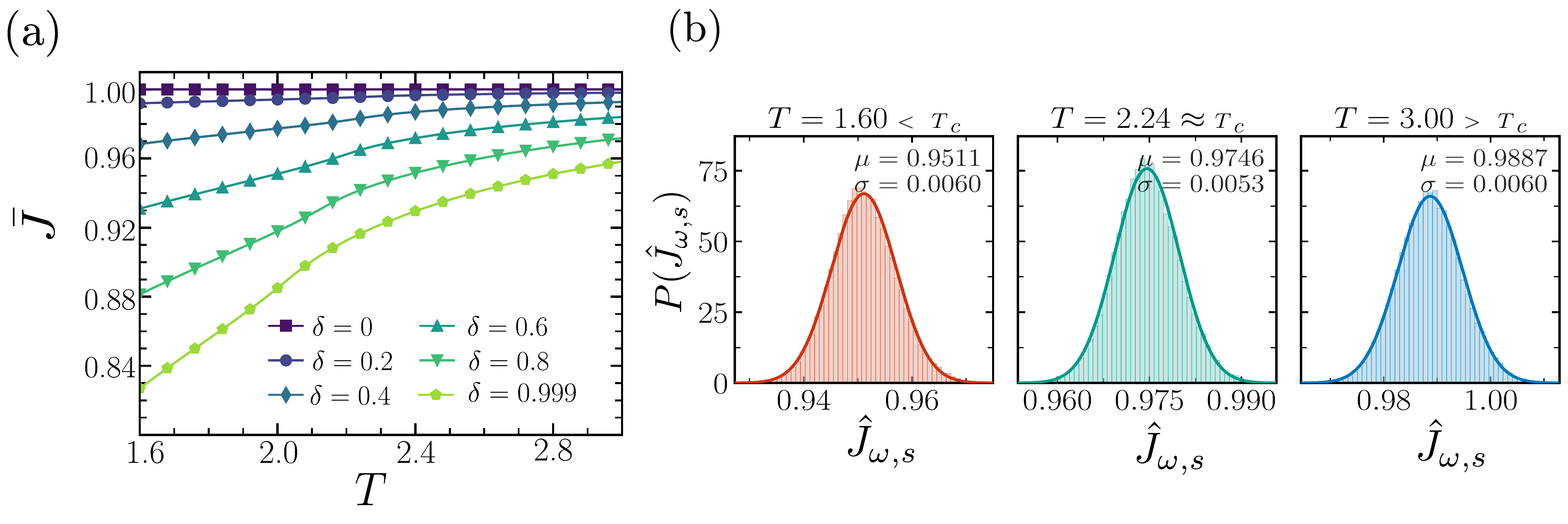}
    \caption{Consensus of inferred couplings.
    \textbf{(a)} Disorder-averaged inferred coupling $\bar{J}$ against temperature for six disorder strengths $\delta \in \{0, 0.2, \dots, 0.999\}$. The clean system recovers $\bar{J} = J_0 = 1$ at every $T$, as it must, since a uniform teacher lies inside the students' hypothesis class. Disorder displaces $\bar{J}$ downward, most strongly in the ordered phase, and the displacement decreases with temperature.
    \textbf{(b)} Consensus distribution $P(\hat{J}_{\omega,s})$ at $\delta = 0.5$ for three temperatures spanning the critical region, with a Gaussian fit (solid) to the full set of $N_\omega \times N_s$ inferred couplings. The mean drifts upward with temperature, from $0.9511$ to $0.9887$, but not reaching $J_0$ within the temperature range used and the width is changing non-monotonically reaching $\sigma = 0.0053$ at $T \approx T_c^{\mathrm{eff}}$ from $\sigma = 0.0060$ both below and above. Learners therefore agree among themselves most closely at criticality but agree most with the truth well above it.}
    \label{fig:2}
\end{figure*}

Fig.~\ref{fig:2} displays both features.  In panel~(a), the clean system recovers the true coupling at all temperatures while every disordered system infers a coupling that falls short of $J_0$. The
shortfall is largest in the ordered phase and shrinks monotonically with temperature. In panel~(b) the width of $P(\hat{J}_{\omega,s})$ at $\delta = 0.5$ changes non-monotonically: it lowers from $\sigma = 0.0060$ at $T = 1.60$ to $\sigma = 0.0053$ at $T \approx T_c^\mathrm{eff}$, then increases back to $\sigma = 0.0060$ at $T = 3.00$. Learners agree with one another most closely at criticality, but agree most closely with the \emph{ground truth} at a temperature $T > T_c^{\mathrm{eff}}$ due to the disorder induced bias. The bias is the displacement of the disorder averaged coupling from the true one, given by
\begin{equation}
\begin{aligned}
    b_{M}(\delta, T)
    &\equiv \mathbb{E}_{\omega,s}\left[\hat{J}_{\omega,s}\right] - J_0, \\
    \lim_{M \to \infty} b_M(\delta,T)
    &\equiv b_{\infty}(\delta,T)
    = \mathbb{E}_{\omega}\left[J_{\omega}^{*}\right] - J_0.
\end{aligned}
\label{eq:27}
\end{equation}
In the infinite-data limit the bias is a misspecification bias, arising from the mismatch between the uniform hypothesis class of the student and the heterogeneous teacher. At finite $M$ the estimator carries, in addition, an estimation bias of $O(1/M)$, arising from its nonlinear dependence on the pattern counts. At $\delta=0$ the mismatch disappears: every realization corresponds to the same uniform teacher, which lies inside the student's hypothesis class, so $b_\infty=0$ at all temperatures and the estimator is consistent for $J_0$. 

To characterize the bias, we expand the inferred coupling $J^*_{\omega}$ by a single student in the infinite data limit around the clean point ($\delta=0$) in order to get 
\begin{equation}
    J^*(\mathbf{J}) = J_0 + g_a \eta_a
    + \tfrac{1}{2} H_{ab}\, \eta_a \eta_b + O(\delta^3),
    \label{eq:28}
\end{equation}
where the gradient $g_a$ and Hessian $H_{ab}$  are evaluated at clean point. A detailed expansion upto $O(\delta^3)$ is shown in Appendix~\ref{sec:appendix_bias}.  For uniform distribution considered, $\langle \eta_a \rangle = 0$ and
$\langle \eta_a \eta_b \rangle = (\delta^2/3)\,\delta_{ab}$, the leading contribution is second order 
\begin{equation}
    b_{\infty}(\delta, T) = \frac{\delta^2}{6} \operatorname{Tr} H(T) + O(\delta^4).
    \label{eq:29}
\end{equation}
Writing $b_{\infty}(\delta, T) = -K(T)\delta^2 + O(\delta^4)$ and using bond symmetry to replace the
trace by $N_b H_{aa}$,
\begin{equation}
    K(T) = -\frac{1}{6} \operatorname{Tr} H(T)
         = -\frac{N_b}{6} H_{aa}(T).
    \label{eq:30}
\end{equation}
An analytical expression for $K(T)$ for finite data is derived in Appendix~%
\ref{app:bias_derivation} for student inference with finite number of samples in terms of the pattern counts $N(x)$ and neighbour sum $S(x)$ as 
\begin{equation}
    K_{loc}(T) = \frac{4}{3T}\,
    \frac{\sum_x N(x)\, S(x)\, \tanh(\beta J_0 S(x))}
         {\sum_x N(x)\, S(x)^2},
    \label{eq:31}
\end{equation}
\begin{figure*}[t!]
    \centering
    \includegraphics[width=0.99\linewidth]{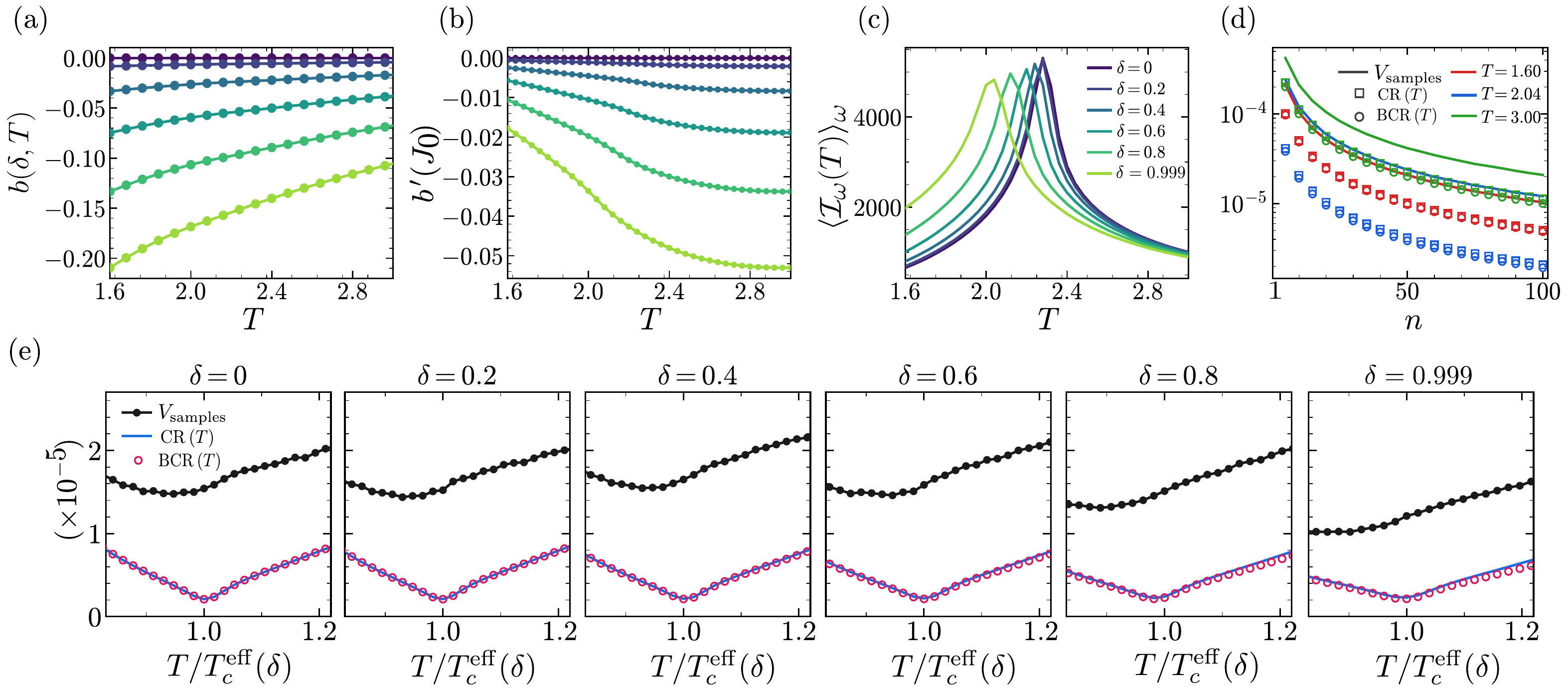}
    \caption{Bias, Fisher information, and the Cram\'er--Rao bound on the
    sampling variance, for six disorder strengths
    $\delta \in \{0, 0.2, 0.4, 0.6, 0.8, 0.999\}$ (colour scale in (c)).
    \textbf{(a)} Measured bias $b_{M}(\delta, T)$ versus temperature using the finite sample expression in Eq.~\eqref{eq:31}. The bias
    vanishes at $\delta = 0$ and grows in magnitude with $\delta$.
    \textbf{(b)} Coupling derivative of the bias $b'(J_0)$, evaluated at $J_0 = 1$. It is negative, small ($|b'| \lesssim 0.05$), and slowly varying, so the prefactor $[1 + b'(J_0)]^2$ in the biased Cram\'er--Rao bound remains close to unity over the whole temperature range.
    \textbf{(c)} Disorder-averaged Fisher information $\mathcal{I_{\omega}}(T)$, which peaks sharply at $T_c^\mathrm{eff}(\delta)$, the specific heat maximum, for every disorder strength.
    \textbf{(d)} Measured sampling variance (solid black) versus sample size $n$ at $\delta = 0.5$, compared with the unbiased (squares) and biased (circles) Cram\'er--Rao bounds at three temperatures $T \in \{1.60, 2.04, 3.00\}$. The measured variance lies above both bounds at every $n$, and the two bounds coincide to within line width because $|b'|$ is small.
    \textbf{(e)} Measured sampling variance (solid black) against the biased
    (red circles) and unbiased (solid blue) Cram\'er--Rao bounds as a function of scaled temperature $T/T_c^\mathrm{eff}(\delta)$, for each disorder strength. The measured variance lies above both bounds at every $\delta$ and reaches its minimum below $T_c^\mathrm{eff}(\delta)$.}
    \label{fig:3}
\end{figure*}

Fig.~\ref{fig:3}(a) shows the measured bias over
$0.8 \leq T/T_c^\mathrm{eff} \leq 1.2$. It vanishes identically at $\delta = 0$,
confirming unbiasedness in the clean limit, and is negative for all
$\delta > 0$, growing in magnitude as $\delta^2$ and decaying with temperature in agreement with Eq.~\eqref{eq:31}. The measured bias is negative, thus $\operatorname{Tr} H(T) < 0$, and hence
$H_{aa} < 0$. And thus shows a downward displacement of the best uniform fit, and thus the student underestimates the bond mean.

At low temperature the spin configurations a student sees are frozen into patterns controlled by the particular bond realization, so the disorder is fully imprinted on the data and the uniform hypothesis class is badly mismatched. As the temperature rises, thermal fluctuations increasingly dominate the configurations, the data become less sensitive to bond-to-bond variation, and the uniform hypothesis becomes an increasingly good description of what the student observes.

\subsection{Sampling variance and the estimator optimum}
\label{subsec:sampling_variance}

Of the two variance components in Eq.~\eqref{eq:26}, only
$V_{\mathrm{samples}}$ is constrained by an information-theoretic bound. At fixed disorder realization $\omega$, all students learn the same teacher and differ only through the finite set of configurations each receives, so their conditional variance
\begin{equation}
    V_{\mathrm{samples},\omega}(T)
    = \operatorname{Var}_s\!\left(\hat{J}_{\omega,s} \mid \omega\right)
    \label{eq:32}
\end{equation}
is a pure finite-sampling quantity, and scales as $1/M$ with the number of
configurations $M$ in a single student's data subset.

To bound Eq.~\eqref{eq:32} one must specify the parametric family within which the bound is taken. For a fixed realization $\omega$, consider the one-parameter family generated by shifting every bond of the true coupling by the same amount,
\begin{equation}
    P_\theta \equiv P_{\bm J^\omega + \epsilon\mathbf{1}},
    \qquad \theta = J_0 + \epsilon .
    \label{eq:33}
\end{equation}
This family contains the true distribution at $\epsilon = 0$ and $\theta$ is the natural scalar parameter for a student that returns a single uniform coupling. Its score is $\partial_\epsilon \ln P = \beta(\mathcal{E} - \langle\mathcal{E}\rangle)$ with
$\mathcal{E} = \sum_{\langle ij\rangle}\sigma_i\sigma_j$, so its Fisher information~\cite{MacKay2004InformationTI} per data is the energy fluctuation~\cite{RevModPhys.67.605,Janke_2004},
\begin{equation}
    \mathcal{I}_\omega(T) = \beta^2\,\mathrm{Var}_{T,\omega}(\mathcal{E}) ,
    \label{eq:34}
\end{equation}
obtained directly from the same configurations the students are trained on.
Since $\mathrm{Var}(\mathcal{E})$ is proportional to the specific heat,
$\mathcal{I}_\omega(T)$ inherits its peak in the critical region: configurations
drawn near criticality are most informative about the effective
coupling~\cite{Mastromatteo_2011}. We define the information-optimal temperature
\begin{equation}
    T_I \equiv \underset{T}{\operatorname{arg\,max}}\;
    \left\langle \mathcal{I}_\omega(T)\right\rangle_\omega .
    \label{eq:35}
\end{equation}

Up to a residual $O(1/M)$ estimation bias, the mean of the student's
estimator within the parametric family is
$\mathbb{E}_s\big(\hat J_{\omega,s}\mid\omega,\epsilon\big)\simeq
J^*(\bm J^\omega+\epsilon\bm 1)$, so its bias is
$b_{\infty,\omega}(\theta)=J^*(\bm J^\omega+\epsilon\bm 1)-\theta$.
The Cram\'er--Rao inequality for a biased estimator~\cite{CR_bound} reads
\begin{equation}
V_{\mathrm{samples},\omega}(T) \;\ge\;
\frac{\Big[\partial_\epsilon\,
\mathbb{E}_s\big(\hat J_{\omega,s}\mid\omega,\epsilon\big)\big|_{\epsilon=0}\Big]^2}
{M\,\mathcal{I}_\omega(T)}
\;\simeq\;
\frac{\big[1+b'_{\infty,\omega}(J_0)\big]^2}{M\,\mathcal{I}_\omega(T)}.
\label{eq:36}
\end{equation}
The derivative is taken with respect to the uniform shift,
\begin{equation}
    b'_{\infty,\omega}(J_0)
    = \left.\frac{\partial b_{\infty,\omega}}{\partial\epsilon}\right|_{\epsilon=0}
    = \sum_{a=1}^{N_b}
      \left.\frac{\partial J^*}{\partial J_a}\right|_{\bm J^\omega} - 1 .
    \label{eq:37}
\end{equation}
Because the shift~\eqref{eq:33} moves the mean coupling at fixed
disorder width, $\partial_\epsilon$ at $\epsilon=0$ is the same
operation as $\partial_{J_0}$ at fixed $T$ and $\delta$, which is the
derivative evaluated in Eq.~\eqref{eq:D19}. The bound and the information are therefore computed in the same family. For $L = 50$ we measure $|b'_{M,\omega}| \lesssim 0.05$ across the whole disorder range with zero bias for $\delta=0$ [Fig.~\ref{fig:3}(b)]. Here the bias derivatives are calculated by taking the total number of samples the teacher has generated.

Averaging Eq.~\eqref{eq:36} over realizations gives the ensemble floor
\begin{equation}
    V_{\mathrm{samples}}(T) \;\geq\; \mathrm{BCR}(T)
    \;=\; \frac{1}{M}\left\langle
    \frac{\left[1+b'_{\infty,\omega}(J_0)\right]^2}{\mathcal{I}_\omega(T)}
    \right\rangle_\omega ,
    \label{eq:38}
\end{equation}

minimized at
\begin{equation}
    T_{\mathrm{CR}} \equiv \underset{T}{\operatorname{arg\,min}}\;
    \mathrm{BCR}(T) .
    \label{eq:39}
\end{equation}
Because the prefactor $[1+b'_{\infty,\omega}]^2$ stays within a few percent of unity and varies slowly with temperature, it does not appreciably displace the minimum, and
\begin{equation}
    T_I \simeq T_{\mathrm{CR}} :
    \label{eq:40}
\end{equation}
the Cram\'er--Rao floor is minimized where the Fisher information peaks. However, the measured sampling variance need not attain its minimum at
$T_{\mathrm{CR}}$, because the students' estimator does not saturate the floor.
Two compressions stand between the data and the inferred coupling: each
configuration is first reduced to its $32$ neighborhood-pattern counts,
discarding the spatial arrangement of the spins, and those counts are then
projected onto a single coupling. We quantify the shortfall by an ensemble
efficiency,
\begin{equation}
    \epsilon(T,\delta) \equiv
    \frac{\mathrm{BCR}(T,\delta)}{V_{\mathrm{samples}}(T,\delta)} ,
    \label{eq:41}
\end{equation}
which is a ratio of disorder-averaged quantities and therefore depends on
disorder as well as on temperature. The estimator is well short of the floor
across the whole window, with $\epsilon$ falling to
$\epsilon \approx 0.15$ at $T_c^{\mathrm{eff}}$ and recovering to
$\epsilon \approx 0.4$ at the edges of the range shown in
Fig.~\ref{fig:3}(e), a substantial loss of
efficiency~\cite{Aurell2012,Hyvrinen2006ConsistencyOP}. Both $\mathrm{BCR}$ and $V_{\mathrm{samples}}$ carry the same $1/M$ prefactor, so the ratio is independent of sample size.

Two properties of $\epsilon$ play different roles: its magnitude sets the
precision, while its slope moves the optimum. Writing
\begin{equation}
    T_s \equiv \underset{T}{\operatorname{arg\,min}}\;
    V_{\mathrm{samples}}(T)
    \label{eq:43}
\end{equation}
for the measured optimum and differentiating
$V_{\mathrm{samples}} = \mathrm{BCR}/\epsilon$ at $T_{\mathrm{CR}}$, where
$\mathrm{BCR}' = 0$ by construction,
\begin{equation}
    \left.\frac{dV_{\mathrm{samples}}}{dT}\right|_{T_{\mathrm{CR}}}
    = -\frac{\mathrm{BCR}(T_{\mathrm{CR}})}{\epsilon(T_{\mathrm{CR}})^2}
    \left.\frac{d\epsilon}{dT}\right|_{T_{\mathrm{CR}}} .
    \label{eq:44}
\end{equation}
The slope of the measured variance at $T_{\mathrm{CR}}$ is thus set entirely by
the slope of the efficiency there, and the two minima coincide only if
$\epsilon'(T_{\mathrm{CR}}) = 0$.

Fig.~\ref{fig:3}(e) shows that $V_{\mathrm{samples}}$ and $\mathrm{BCR}$ differ not only in magnitude but in shape. The bound has a minimum at $T_{\mathrm{CR}} \simeq T_c^{\mathrm{eff}}$, inherited from the peak of the Fisher information, whereas the measured variance varies weakly over the same window and its minimum is displaced downward,
\begin{equation}
    T_s < T_{\mathrm{CR}} \qquad\text{for every } \delta \in [0,1),
    \label{eq:45}
\end{equation}
with the displacement growing with disorder, from $\approx 0.03$ at $\delta = 0$ to $\approx 0.28$ at $\delta = 0.999$ for $L = 50$. Equivalently, by
Eq.~\eqref{eq:44}, $\epsilon$ is falling at $T_{\mathrm{CR}}$: the projection loses efficiency fastest where the bound is tightest, so the critical enhancement of the information is only partly converted into precision by an estimator that discards the spatial arrangement of the spins. The two temperatures are set by competing trends: as $T$ falls below $T_{\mathrm{CR}}$ the data become less informative, since the Fisher information falls away from its peak, while the efficiency of the projection improves. The measured optimum $T_s$ is where the
two balance. The growth of the displacement with $\delta$ has two sources, which we do not attempt to separate here: the efficiency slope
$\epsilon'(T_{\mathrm{CR}})$ itself varies with disorder, and the
Fisher-information peak broadens [Fig.~\ref{fig:3}(c)], so that a given
efficiency slope displaces the minimum further when the $\mathrm{BCR}$ minimum is flatter.

Consensus therefore selects a well-defined estimator optimum $T_s(\delta)$,
displaced into the ordered region from $T_I \simeq T_{\mathrm{CR}}$. Thus, within a teacher, students agree most closely at a temperature below the one at which their data are most informative. This displacement is a property of the particular
$N(x)$-weighted projection used here rather than a general feature of inference on this model; the information itself remains concentrated in the critical region.

\subsection{Disorder variance}
\label{subsec:disorder_variance}

The second component of the consensus width is the variance across teachers,
\begin{equation}
    V_{\mathrm{disorder}}
    = \operatorname{Var}_\omega\!\left[\mathbb{E}_s\!\left(\hat{J}_{\omega,s} \mid \omega\right)\right]
    = \operatorname{Var}_\omega\!\left(\bar{J}_\omega\right),
    \label{eq:46}
\end{equation}
where $\bar{J}_\omega$ is the inferred couplings averaged over the
$N_s$ students that share the realization $\omega$. 

It is useful to separate this description of the $V_{\mathrm{disorder}}$ from its infinite-data limit. If each teacher were assigned
a single student receiving the full teacher distribution, that student would infer $J^*_\omega$ and the variance between realizations would be
\begin{equation}
    V_{\mathrm{disorder}}^{\infty}
    = \operatorname{Var}_\omega\!\left(J^*_\omega\right).
    \label{eq:47}
\end{equation}
Equations~\eqref{eq:46} and \eqref{eq:47} coincide only as $M \to \infty$;
at finite $M$ they differ because $\mathbb{E}_s(\hat{J} \mid \omega)$ inherits a residual $O(1/M)$ dependence on the sampling. All of the analysis below is carried out for Eq.~\eqref{eq:47}, which is the object accessible to the disorder expansion.

\begin{figure*}[t!]
    \centering
    \includegraphics[width=0.99\linewidth]{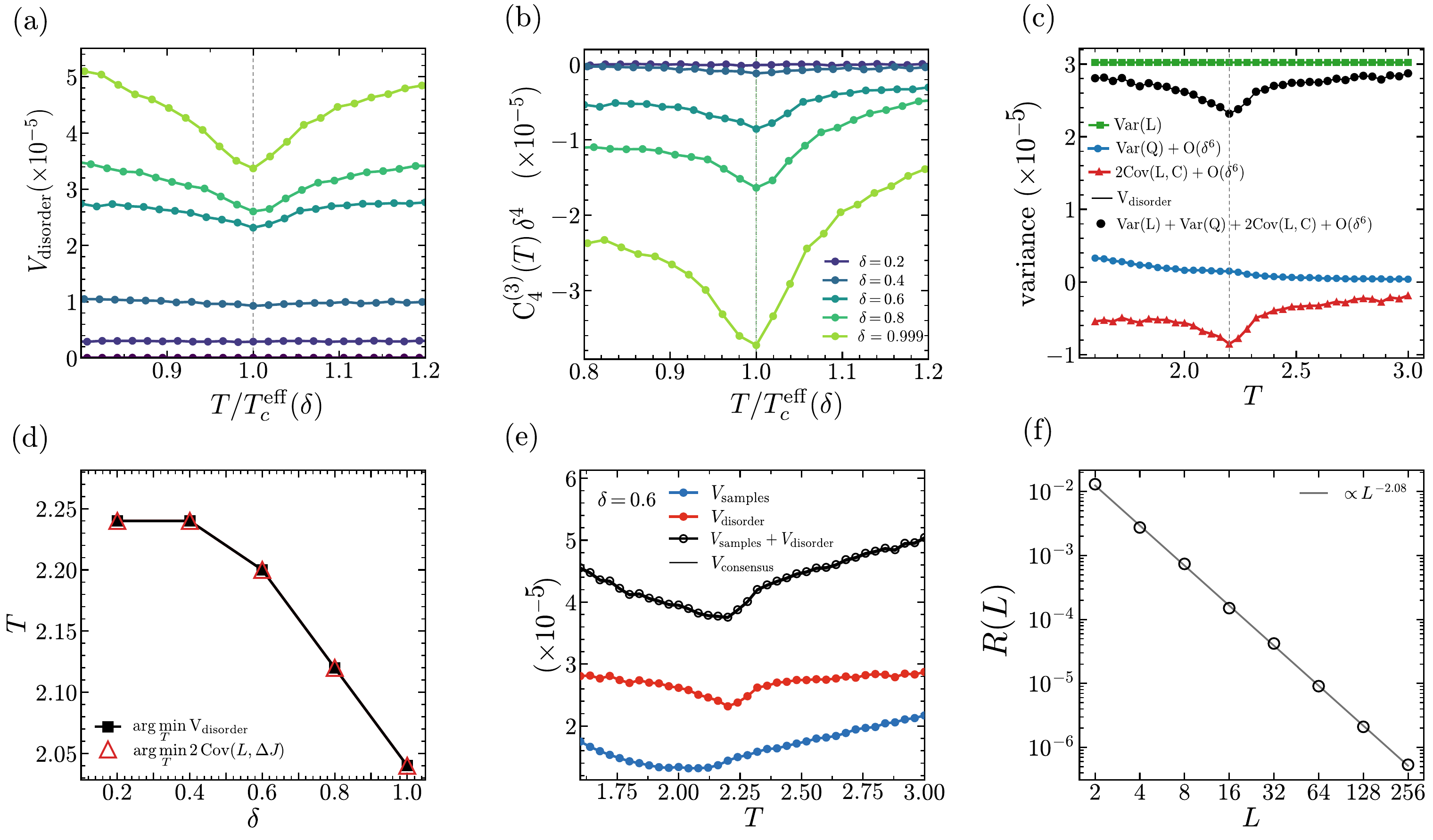}
    \caption{Origin of the critical minimum in the disorder variance. Throughout,
$L$, $Q$ and $C$ denote the linear, quadratic and cubic responses of
Eq.~\eqref{eq:48}, $L = g_a\eta_a$, $Q = \tfrac{1}{2}H_{ab}\eta_a\eta_b$ and
$C = \tfrac{1}{6}\Theta_{abc}\eta_a\eta_b\eta_c$, with $\eta_a$ the offset of
bond $a$ from $J_0$.
    \textbf{(a)} $V_{\mathrm{disorder}}$ versus scaled temperature for
    $\delta \in \{0.2, \dots, 0.999\}$ (color scale used in (b)). The curves are flat at weak disorder and develop a minimum at $T/T_c^{\mathrm{eff}}(\delta)$ that deepens with $\delta$. Values of $T_c^{\mathrm{eff}}(\delta)$ used to scale the axis are the corresponding maxima of specific heat.
    \textbf{(b)} The linear-cubic covariance $2\operatorname{Cov}(L, C) = C_4^{(3)}(T)\delta^4$, for the same disorder strengths. It is negative throughout, most negative at $T/T_c^{\mathrm{eff}}(\delta) = 0.999$.
    \textbf{(c)} Term-by-term evaluation of Eq.~\eqref{eq:50} at $\delta = 0.6$.
    $\operatorname{Var}(L)$ (green) is temperature-independent floor, since $g_a = 1/N_b$ exactly. $\operatorname{Var}(Q)$ (blue) is monotonic; only $2\operatorname{Cov}(L, C)$ (red) has
    an extremum. Their sum (black dots) reproduces the measured $V_{\mathrm{disorder}}$ (black line).
    \textbf{(d)} The temperature minimizing $V_{\mathrm{disorder}}$ (squares) and the temperature minimizing $2\operatorname{Cov}(L, C)$ (triangles), each calculated from $N_\omega=200$ realizations.
    \textbf{(e)} Additivity of the consensus width at $\delta = 0.6$:
    $V_{\mathrm{samples}} + V_{\mathrm{disorder}}$ (open circles) against the measured $V_{\mathrm{consensus}}$ (line).
    \textbf{(f)} Finite-size scaling of the self-averaging ratio
    $R(L)=
        \frac{V_{\mathrm{disorder}}}
        {\left[\mathbb{E}_{\omega}(J_\omega^*)\right]^2}$,
    which decays approximately as $L^{-2.08}$, consistent with the leading
    $1/L^2$ behavior.
    \label{fig:4}
    }
\end{figure*}

Figure~\ref{fig:4}(a) shows the behavior to be explained. At weak disorder $V_{\mathrm{disorder}}$ is essentially flat in temperature. As $\delta$ grows, a minimum develops, and it sits at $T/T_c^{\mathrm{eff}}(\delta) = 0.999$ rather than being displaced to either side. The disorder variance is thus smallest exactly where the data are most strongly correlated. Two features need explaining: a temperature-independent floor, and the critical minimum that develops on top of it. They arise at different orders of the same expansion.

Continuing the expansion of Eq.~\eqref{eq:28} to higher order gives
\begin{equation}
    J^*(\mathbf{J}) = J_0 + g_a \eta_a
    + \tfrac{1}{2} H_{ab}\, \eta_a \eta_b
    + \tfrac{1}{6} \Theta_{abc}\, \eta_a \eta_b \eta_c + O(\delta^4),
    \label{eq:48}
\end{equation}
where $g_a$, $H_{ab}$ and $\Theta_{abc}$ are the first three bond derivatives of $J^*$ evaluated at the clean point, and $\eta_a$ is the offset of bond $a$ from $J_0$. The linear coefficient is a constant, $g_a = 1/N_b$ with $N_b = 2L^2$ (See Eq.~\eqref{eq:C4} in Appendix~\ref{sec:appendix_bias}). On writing
\begin{equation}
    L = g_a \eta_a, \qquad
    Q = \tfrac{1}{2} H_{ab} \eta_a \eta_b, \qquad
    C = \tfrac{1}{6} \Theta_{abc} \eta_a \eta_b \eta_c,
    \label{eq:49}
\end{equation}
and taking the variance of Eq.~\eqref{eq:48} about $J_0$,
\begin{equation}
    V_{\mathrm{disorder}}
    = \operatorname{Var}(L) + \operatorname{Var}(Q)
      + 2\operatorname{Cov}(L, C) + O(\delta^6).
    \label{eq:50}
\end{equation}
Only three terms survive at this order: $\operatorname{Cov}(L, Q)$ is odd in $\eta$ and
vanishes by the symmetry of the disorder distribution, while $\operatorname{Var}(C)$ and
$\operatorname{Cov}(Q, C)$ are $O(\delta^6)$ and $O(\delta^5)$ respectively, the latter also
vanishing by symmetry. Evaluating the moments gives
\begin{equation}
    V_{\mathrm{disorder}}(T, \delta, L)
    = \frac{\delta^2}{3N_b} + \delta^4 C_4(T, L) + O(\delta^6),
    \label{eq:51}
\end{equation}
with $C_4(T,L) = C_4^{(H)}(T,L) + C_4^{(3)}(T,L)$ and
\begin{equation}
    C_4^{(H)} = \frac{1}{18}\operatorname{Tr}(H^2)
    - \frac{1}{30}\sum_a H_{aa}^2 ,
    \label{eq:52}
\end{equation}
\begin{equation}
    C_4^{(3)} = \frac{1}{9N_b}\sum_{a,c} \Theta_{acc}
    - \frac{2}{45 N_b}\sum_a \Theta_{aaa} .
    \label{eq:53}
\end{equation}
A detailed derivation is given in Appendix~\ref{sec:disorder_variance_appendix}.

The leading term of Eq.~\eqref{eq:51} is $O(\delta^2)$ and temperature
independent. It is strongly self-averaging, vanishing as $1/L^2$, which is the scaling confirmed by the $L^{-2.08}$ decay of the self-averaging ratio
$R(L) = V_{\mathrm{disorder}}/[\mathbb{E}_\omega(\bar{J}_\omega)]^2$~\cite{Aharony1996} in Fig.~\ref{fig:4}(f). The critical minimum therefore cannot come from linear response, and the question becomes at which order in $\delta$ the temperature first enters.

The two quartic coefficients differ in an essential way. $C_4^{(H)}$ is non-negative, since $\delta^4 C_4^{(H)} = \operatorname{Var}(Q)$ is a variance. By contrast $C_4^{(3)}$ is a covariance and carries no definite sign. This distinction is what decides which term produces the minimum.

Figure~\ref{fig:4}(c) evaluates the three variance terms of Eq.~\eqref{eq:50} separately at $\delta = 0.6$, plotted against $T$. The linear variance (green) is a constant: with $g_a = 1/N_b$ exactly. Its value is $\delta^2/3N_b = 2.40\times10^{-5}$ at $M,N_\omega \rightarrow\infty$ for $\delta = 0.6$ and for $L = 50$ with $N_\omega=200$ and $M =100$ it comes out as $3\times10^{-5}$. The measured floor scatters on both sides of the analytic value across disorder strengths, with ratios $1.09$, $0.96$, $1.26$, $0.88$ and $0.88$ at $\delta = 0.2$, $0.4$, $0.6$, $0.8$ and $0.999$. The Hessian variance is positive and temperature dependent, but monotonic, with no extremum. Only $2\operatorname{Cov}(L, C)$ is non-monotonic: it is negative throughout and reaches its most negative value near $T_c^{\mathrm{eff}}$. Summing the two variances and the covariance reproduces the independently measured $V_{\mathrm{disorder}}$ across the whole range, confirming that the expansion is converged at $O(\delta^4)$ for this disorder strength.

Fig.~\ref{fig:4}(b) shows the covariance for all disorder strengths. It is negative at every temperature and dips at $T/T_c^{\mathrm{eff}}(\delta) = 0.999$, with the dip deepening as $\delta$ grows. Fig.~\ref{fig:4}(d) shows that the temperature minimizing $V_{\mathrm{disorder}}$ and the temperature minimizing $2\operatorname{Cov}(L, C)$ agree at every $\delta$ within their fitted uncertainties, and that both drift downward together from $T \approx 2.24$ to $T \approx 2.04$ with increasing disorder, tracking $T_c^{\mathrm{eff}}(\delta)$. The minimum of the disorder variance is therefore the minimum of the linear-cubic covariance, it is the negative covariance between the linear and cubic responses, not either variance on its own, that produces the critical minimum.

It remains to explain how $H_{ab}$ and $\Theta_{abc}$ depend on temperature, and how they behave as $T \to \infty$. The scaling form derived in Appendix~\ref{sec:scaling_form_appendix} gives,
\begin{equation}
    g_a = \Phi'_a, \qquad
    H_{ab} = \beta\, \Phi''_{ab}, \qquad
    \Theta_{abc} = \beta^2\, \Phi'''_{abc},
    \label{eq:54}
\end{equation}
where $\Phi$ is the scaling function which can be written from Eq.~\eqref{eq:16} (See Appendix~\ref{sec:scaling_form_appendix}). In particular $g_a$ carries no $\beta$ dependence, consistent with it being the constant $1/N_b$ established above. Since Hessian depends on $\beta$,  the bias $b = (\delta^2/6)\operatorname{Tr}H \propto \beta$, the bias falls off as $1/T$ as in Eq.~\eqref{eq:31}.

In the $T \to \infty$ limit $H, \Theta \to 0$, so $C_4 \to 0$ and
Eq.~\eqref{eq:51} reduces to the bare floor $\delta^2/3N_b$. Equivalently,
Eq.~\eqref{eq:48} becomes an identity and the student simply
recovers the arithmetic mean of its teacher's bonds.

Finally, Fig.~\ref{fig:4}(e) verifies that the two components combine
additively, $V_{\mathrm{consensus}} = V_{\mathrm{samples}} + V_{\mathrm{disorder}}$, at
$\delta = 0.6$. Since both components are individually minimized near $T_c^{\mathrm{eff}}$,
the consensus width inherits the critical minimum observed in
Fig.~\ref{fig:2}(b).

\subsection{Total consensus variance}
\label{subsec:total_variance}

$V_{\mathrm{samples}}$ is bounded below by the inverse Fisher information, which peaks at $T_c^{\mathrm{eff}}$. $V_{\mathrm{disorder}}$ is suppressed there because the cubic disorder response is anticorrelated with the linear disorder response, narrowing the spread of couplings most strongly at criticality. Fig.~\ref{fig:4}(e) confirms that the decomposition holds numerically at $\delta = 0.6$; we have verified it across the full range of disorder strengths and sample sizes.

Fig.~\ref{fig:5}(a) shows the measured consensus width
$\mathcal{W} = \sqrt{V_{\mathrm{consensus}}}$ against scaled temperature. A
minimum appears near $T_c^{\mathrm{eff}}(\delta)$ for every disorder strength, and the minima sharpens and gets pins down exactly at $T_c^{\mathrm{eff}}(\delta)$ as $\delta$ grows showing that disorder does not wash out the critical signature in consensus, instead it sharpens it.

At small $\delta$ the disorder variance is suppressed as $\delta^2$ and the width is set almost entirely by sampling component. At large $\delta$ the disorder variance grows and eventually dominates, leaving the sampling contribution as a small correction. The crossover between the two depends on $M$, the number of samples each student receives.

Combining the Cram\'er--Rao inequality with the decomposition gives,
\begin{equation}
    V_{\mathrm{consensus}}(T) \;\geq\; \mathrm{BCR}(T) + V_{\mathrm{disorder}}(T),
    \label{eq:55}
\end{equation}
The first is reducible: it falls as $1/M$, so any learner can approach it by collecting more configurations, though no inference procedure can go below it. The second is irreducible at fixed $L$ and $\delta$. It vanishes
only in the limits where that spread disappears: at $\delta = 0$, where all
teachers coincide, and as $L \to \infty$, where the $\delta^2/6L^2$ floor
self-averages away (Fig.~\ref{fig:4}(f)). In either limit the consensus variance reduces to the sampling term alone, and is then bounded below by a quantity that itself has a sharp minimum at the transition.

It is worth being precise about where each minimum sits in the two mechanisms. The Cram\'er--Rao floor is minimized at $T_{\mathrm{CR}}$, but the estimator efficiency displaces the measured minimum to $T_s < T_{\mathrm{CR}}$. In the disorder sector the leading $O(\delta^2)$ term is temperature independent, so all temperature dependence enters through the nonlinear response at $O(\delta^4)$; writing
\begin{equation}
    T_d \equiv \underset{T}{\operatorname{arg\,min}}\;
    V_{\mathrm{disorder}}(T),
    \label{eq:56}
\end{equation}
we find $T_d$ to coincide with the pseudocritical temperature. Thus the total consensus width remains smallest near the critical region for every disorder strength.

\subsection{Mean squared error and the optimal temperature for inference}
\label{sec:mse}

\begin{figure*}[t!]
   
    \includegraphics[width=0.99\linewidth]{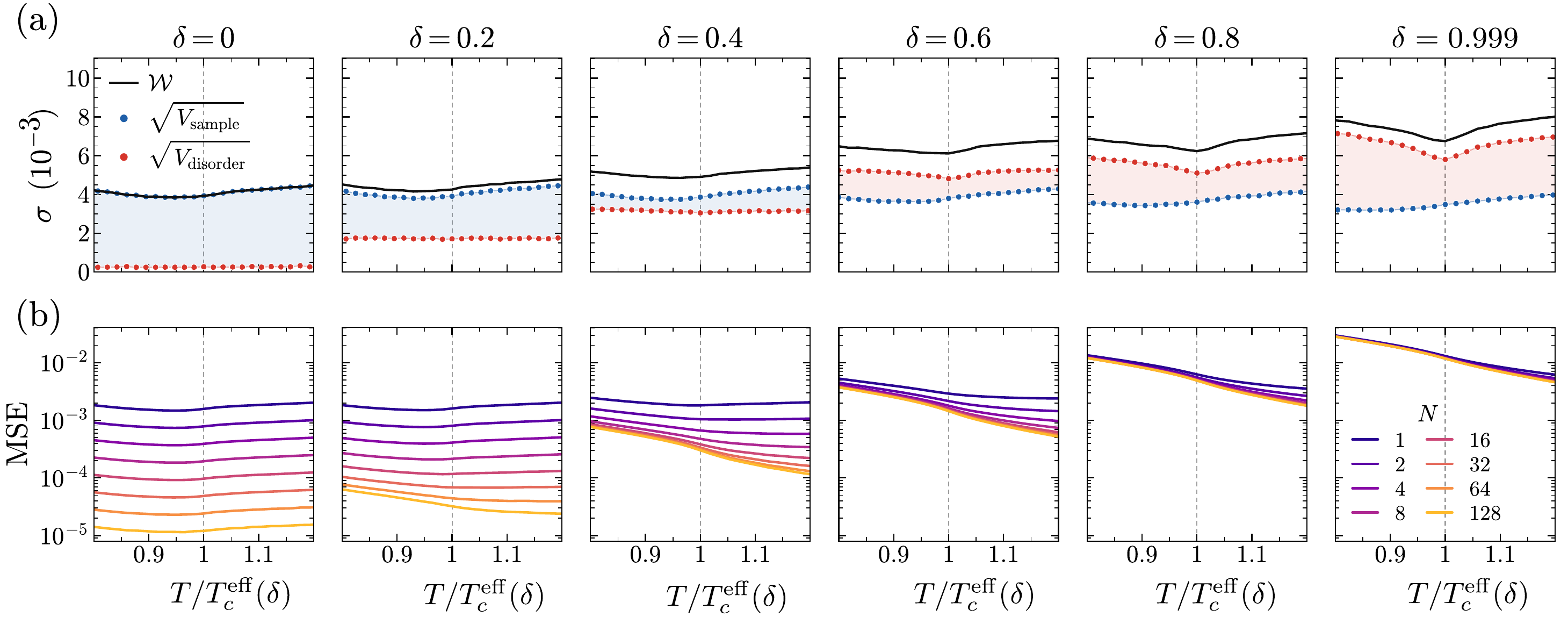}
    \caption{Consensus width and mean squared error across disorder strengths.
    \textbf{(a)} Consensus width
    $\mathcal W=\sqrt{V_\mathrm{consensus}}$ as a function
    of scaled temperature $T/T_c^{\mathrm{eff}}(\delta)$, decomposed into sampling and disorder contributions. The total width is smallest in the critical region. The sampling contribution has a nearby, slightly shifted minimum because the projected estimator has temperature-dependent efficiency, while the disorder contribution is minimized at $T_c$ and acquires its temperature dependence through nonlinear disorder response.
    \textbf{(b)} Mean squared error $\mathrm{MSE}(\hat J)$ for the same
    disorder strengths and several per-student sample sizes $N$. Increasing
    $N$ reduces the variance contribution but not the disorder-induced
    bias floor. At stronger disorder means stronger bias which masks the critical minimum that remains visible in the consensus width.}
    \label{fig:5}
\end{figure*}

Everything so far has measured agreement rather than accuracy. We now ask how
close to the truth the inference actually is, through the mean squared error,
\begin{equation}
    \mathrm{MSE}(\hat J; T)
    = b_M(\delta,T)^2 + V_{\rm samples}(T) + V_{\rm disorder}(T),
    \label{eq:57}
\end{equation}
where $b_M$ and $V_{\rm disorder}$ are the bias and disorder variance
measured with a finite number of samples per student. In the large-$M$ limit both
approach their population values,
\begin{equation}
    b_M \to b_\infty = \mathbb{E}_\omega[J^*_\omega] - J_0,
    \qquad
    V_{\rm disorder} \to V^{\infty}_{\rm disorder}
    = \operatorname{Var}_\omega(J^*_\omega),
    \label{eq:58}
\end{equation}
The MSE therefore differs from the consensus variance by the squared bias. Let $T_V \equiv \arg\min_T V_{\rm consensus}(T)$ denote the temperature of
tightest consensus, and
\begin{equation}
    T_{\rm opt} \equiv \arg\min_T \mathrm{MSE}(\hat J; T)
    \label{eq:59}
\end{equation}
the temperature of best reconstruction of the inferred parameter. Using $V_{\rm consensus} = V_{\rm samples} + V_{\rm disorder}$ and
differentiating Eq.~\eqref{eq:57} at $T_V$, only the bias survives,
\begin{equation}
    \left.\frac{d\,\mathrm{MSE}}{dT}\right|_{T_V}
    = 2\,b_M(T_V)\,\partial_T b_M(T_V) < 0
    \label{eq:60}
\end{equation}
The sign following because the bias is negative and increases monotonically
toward zero with temperature, vanishing only as $T\to\infty$. The MSE is thus
still decreasing where the consensus variance has already reached its minimum, so
\begin{equation}
    T_{\rm opt} > T_V .
    \label{eq:61}
\end{equation}
When the bias is weak the displacement can be quantified. Expanding the two
variances about their minima $T_s$ and $T_d$, where their first derivatives
vanish, and the squared bias about $T_d$, where it is at no extremum and so
varies linearly,
\begin{align}
    V_{\rm samples}(T) &\simeq V_s^{\min}
    + \tfrac12\kappa_s(T-T_s)^2,\\
    V_{\rm disorder}(T) &\simeq V_d^{\min}
    + \tfrac12\kappa_d(T-T_d)^2,\\
    b_M(T)^2 &\simeq b_{M,d}^2 + q_d\,(T-T_d),
\end{align}
with $\kappa_s = V_{\rm samples}''(T_s) > 0$ and
$\kappa_d = V_{\rm disorder}''(T_d) > 0$, $b_{M,d} \equiv b_M(T_d)$ and $q_d \equiv 2b_{M,d}\,\partial_T b_M|_{T_d} < 0$. Setting $\mathrm{MSE}'(T_{\rm opt}) = 0$ gives
\begin{equation}
    T_{\rm opt} \simeq
    \frac{\kappa_s T_s + \kappa_d T_d - q_d}{\kappa_s + \kappa_d},
    \label{eq:65}
\end{equation}
Applying the same expansion to $V_{\rm consensus}$ alone places its minimum at $T_V = (\kappa_sT_s + \kappa_dT_d)/(\kappa_s+\kappa_d)$. Equation
\eqref{eq:65} is therefore
\begin{equation}
    T_{\rm opt} \simeq T_V + \frac{|q_d|}{\kappa_s + \kappa_d},
    \label{eq:66}
\end{equation}
which is the quantitative form of Eq.~\eqref{eq:61}: the bias
displaces the reconstruction optimum above the consensus optimum by an amount set by the ratio of the bias slope to the combined curvature of the two variances.

If the bias is negligible across the window, $T_{\rm opt} \simeq T_V$ and the optimum lies between $T_s$ and $T_d$; with an efficient estimator $T_s \simeq T_{\rm CR}$ and, since $T_d \simeq T_c^{\rm eff}$, best inference then occurs in the critical region. As the bias grows, $|q_d|$ pushes $T_{\rm opt}$ upward, and once the displacement exceeds $T_d - T_V$ the optimum leaves the critical region altogether.

Fig.~\ref{fig:5}(b) shows exactly this. At $\delta = 0$ there is no bias and no disorder variance, so $\mathrm{MSE} = V_{\rm samples}$ and the MSE minimum coincides with the consensus minimum for every $M$. As $\delta$ increases the two separate, with the MSE minimum displaced to higher temperature as Eq.~\eqref{eq:66} requires. Beyond $\delta \simeq 0.4$ the squared bias dominates, the MSE becomes monotonically decreasing across the window, and the optimum has left the critical region entirely. The same crossover occurs as $M$ grows at fixed $\delta$, since more data suppresses $V_{\rm samples}$ but leaves the bias floor untouched. Throughout, the critical minimum in the consensus width
[Fig.~\ref{fig:5}(a)] is unaffected.

\section{Conclusion and Discussion}
\label{sec:conclusion}

We have introduced consensus as a measure of the reproducibility of
effective-model inference in a disordered statistical system. In the random-bond Ising model the consensus variance separates exactly into finite-sampling and quenched-disorder contributions.

In the sampling sector the Fisher information defines an information optimum temperature $T_I$ in the critical region, and the biased Cram\'er--Rao floor is minimized at a nearby temperature $T_{\rm CR}$, the two coinciding because the bias derivative is negligible. The measured sampling variance attains its minimum at $T_s < T_{\rm CR}$. This displacement is a property of the estimator and the information itself remains unambiguously concentrated in the critical region.

The disorder variance is controlled by a different mechanism. The dominating term $O(\delta^2)$ in the expansion of disorder variance is a temperature independent floor. Temperature enters first at fourth order through the variance of the Hessian and the covariance between the first and third derivative. It was  found that the covariance term is the origin of the critical minimum at $T_d$. 

The total consensus variance have components whose minimum lies at $T_s$ and $T_d$. $T_d$ is found to be pseudocritical temperature $T_c^{\mathrm{eff}}(\delta)$ for disorder $\delta$. But $T_s$ lies short of $T_c^{\mathrm{eff}}(\delta)$ . When disorder $\delta=0$, the region of maximum consensus is the critical regime when the estimator is efficient. But as the disorder grows stronger, disorder variance dominates the sampling term and pins the consensus minima more sharply at criticality irrespective of the efficiency of the estimator.

These two variances, together with the bias, determine the temperature of best reconstruction. Because the bias is negative and decays monotonically with temperature, the mean squared error is still falling where the consensus variance has already reached its minimum. So, the optimal temperature for inference, $T_{\rm opt}$ always lies above the consensus optimum when the bias is appreciable. When the bias is weak (Eq.~\eqref{eq:66}), it places the optimum between the two variance minima,
$T_s < T_{\rm opt} < T_d$, as the bias grow stronger, $T_{\rm opt}$
leaves the interval and moves into the high-temperature regime. The information optimum lies in the critical region, while the reconstruction optimum is selected jointly by estimator efficiency, variance and bias.

For a clean system inferred with an efficient estimator, there is no bias and no disorder variance, and accuracy is limited purely by sampling noise; the optimum then sits at the information maximum, in the critical region. Criticality retains the role whenever sampling noise dominates with limited data and weak disorder, because the critical enhancement of the Fisher information is precisely what compensates for having few configurations to learn from \cite{Mastromatteo_2011}. In the other end, with abundant data and strong disorder, the sampling variance has been driven down while the bias floor remains untouched, the misspecification error dominates the mean squared error, and the optimum moves away from criticality into the high-temperature regime. Strong disorder therefore makes accurate inference hard at any finite
temperature. This suggests that estimating and correcting the bias from the samples themselves would be valuable: if the misspecification bias can be removed, the optimum returns to the critical region, where the enhanced Fisher information allows accurate inference from far fewer configurations.

Consensus is therefore not a relabeling of a variance decomposition. By
separating reproducibility from accuracy it makes explicit which part of
inference is information limited, which part is estimator dependent, and which part arises from quenched heterogeneity or model misspecification. Nothing in the construction is specific to the Ising model: the same decomposition applies wherever a heterogeneous system is described by a coarse-grained effective model inferred from finite data~\cite{Machta2013}, and we expect the separation between the information optimum and the reconstruction optimum to be generic.

\section*{Acknowledgments}

I am grateful to S.~S. Ashwin, who proposed the problem. Part of this work was presented in the Spring College on the Physics of Complex Systems, ICTP Trieste, February-March 2026, where Matteo Marsili suggested the Cram\'er-Rao analysis of the sampling variance. I thank ICTP for financial support enabling my participation. Computations were carried out on the
high-performance computing facility at GITAM University, Bengaluru.

\appendix

\section{Monte Carlo sampling protocol}
\label{app:MC}

Configurations were generated by Markov chain Monte Carlo on the $L\times L$
square lattice with periodic boundary conditions, using two update schemes across the temperature ranges. For $1.60\leq T\leq 2.40$ we used Wolff
single-cluster updates~\cite{Wolff}. Since $J_{ij}\sim\mathcal{U}(J_0-\delta,J_0+\delta)$ with $J_0=1$ and $\delta<1$, all couplings are non-negative and the construction remains valid throughout. For $2.40<T\leq3.00$ we used single spin flip Metropolis updates~\cite{metropolis}, as smaller clusters on higher temperature make Wolff algorithm inefficient there. The crossover temperature $T=2.40$ lies above $T_c^{\rm eff}(\delta)$ for every disorder strength, so the entire critical region is covered by the Wolff cluster algorithm.

The quenched bonds were generated once per realization and held fixed at every temperature, so all temperature dependence reported here is measured at fixed disorder. Each run was initialized randomly and equilibrated according to the system size and sampling protocol. For example, for $L=50$, near  $T_c^{\rm eff}(\delta=0.5)$, $3\times10^4$ cluster updates was done for equilibration. The equilibrated samples are stored every $10\tau$ intervals with the $\tau$ being largest near critical temperature and increases with system size $L$ and disorder $\delta$ value.

Realizations were run as independent jobs in the range $T\in[1.60,3.00]$ in
steps of $\Delta T=0.04$, with $N_\omega=200$ realizations per disorder
strength. In the consensus measurements, each teacher has $N_s=250$ students receiving $M=100$ configurations each. Thus a total of $N_{\mathrm{tot}} =25000$ independent equilibrium configurations has been generated per $(\omega,T)$.

\section{The sixteen neighbourhood patterns}
\label{sec:patterns}

The nearest-neighbour Ising model has the property that the conditional probability  of a single spin depends on the rest of the lattice only through its four neighbours,
\begin{equation}
    P_\omega(\sigma_i = +1 \mid \{\sigma_j\}) = \frac{1}{1+e^{-2h_i}},
    \qquad
    h_i = \beta \sum_{j \in \partial i} J_{ij}\sigma_j .
    \label{eq:B1}
\end{equation}

Since each neighbour takes two values, the neighbourhood is specified by four binary digits, giving $2^4 = 16$ distinct patterns.

We encode them in binary, mapping each spin to a single bit,
\begin{equation}
    b = \tfrac{1}{2}(\sigma + 1) =
    \begin{cases}
        1, & \sigma = +1,\\
        0, & \sigma = -1.
    \end{cases}
    \label{eq:B2}
\end{equation}

and assigning the four neighbours the place values $8, 4, 2, 1$ in the order
up, down, left, right:
\begin{equation}
    x = 8\,b_{\rm up} + 4\,b_{\rm down} + 2\,b_{\rm left} + b_{\rm right}
    \;\in\; \{0, 1, \dots, 15\}.
    \label{eq:B3}
\end{equation}

Table~\ref{tab:patterns} lists all sixteen patterns, together with the
neighbour sum $S(x)$ and the probability $P(\sigma_i{=}+1\mid x)$ that the
central spin is up. Scanning a stored configuration site by site and incrementing a counter according to the pattern $x$ and the sign of the central spin yields the pair
\begin{equation}
    n_+(x),\; n_-(x),
    \qquad
    N(x) = n_+(x) + n_-(x),
    \label{eq:B4}
\end{equation}  

The configuration is therefore stored as the $32$-component vector
\begin{equation}
    \bigl(n_+(0), n_-(0), n_+(1), n_-(1), \dots, n_+(15), n_-(15)\bigr),
    \label{eq:B5}
\end{equation}
in place of the $L^2$ spins.

\begin{table}[t]
\centering
\caption{The sixteen neighbourhood patterns. Bits are ordered
$(b_{\rm up}, b_{\rm down}, b_{\rm left}, b_{\rm right})$ with place values
$8,4,2,1$. The last column is the exact conditional probability
$P(\sigma_i{=}+1\mid x) = [1+e^{-2\beta J_0 S(x)}]^{-1}$ for a clean teacher with
$J_0 = 1$ at $T = 2.269$.}
\label{tab:patterns}
\small
\begin{tabular}{cccrc}
\hline\hline
$x$ & bits & $(\sigma_{\rm u},\sigma_{\rm d},\sigma_{\rm l},\sigma_{\rm r})$
    & $S(x)$ & $P(\sigma_i=+1\mid x)$ \\
\hline
 0 & 0000 & $(-,-,-,-)$ & $-4$ & $0.0286$ \\
 1 & 0001 & $(-,-,-,+)$ & $-2$ & $0.1464$ \\
 2 & 0010 & $(-,-,+,-)$ & $-2$ & $0.1464$ \\
 4 & 0100 & $(-,+,-,-)$ & $-2$ & $0.1464$ \\
 8 & 1000 & $(+,-,-,-)$ & $-2$ & $0.1464$ \\
 3 & 0011 & $(-,-,+,+)$ & $0$  & $0.5000$ \\
 5 & 0101 & $(-,+,-,+)$ & $0$  & $0.5000$ \\
 6 & 0110 & $(-,+,+,-)$ & $0$  & $0.5000$ \\
 9 & 1001 & $(+,-,-,+)$ & $0$  & $0.5000$ \\
10 & 1010 & $(+,-,+,-)$ & $0$  & $0.5000$ \\
12 & 1100 & $(+,+,-,-)$ & $0$  & $0.5000$ \\
 7 & 0111 & $(-,+,+,+)$ & $+2$ & $0.8536$ \\
11 & 1011 & $(+,-,+,+)$ & $+2$ & $0.8536$ \\
13 & 1101 & $(+,+,-,+)$ & $+2$ & $0.8536$ \\
14 & 1110 & $(+,+,+,-)$ & $+2$ & $0.8536$ \\
15 & 1111 & $(+,+,+,+)$ & $+4$ & $0.9714$ \\
\hline\hline
\end{tabular}
\end{table}

\section{Weak-disorder expansion}
\label{app:moments}

Let $J^*(\bm J)$ be the coupling recovered by an ideal student in the
infinite-data limit from a teacher with bond configuration
$\bm J = (J_1,\dots,J_{N_b}) \in \mathbb{R}^{N_b}$, where $N_b = 2L^2$. Writing
$J_a = J_0 + \eta_a$, so that $\bm J = J_0\mathbf{1} + \bm\eta$, and expanding
about the clean point,
\begin{equation}
    J^*(\bm J) = J_0 + g_a\eta_a + \tfrac12 H_{ab}\eta_a\eta_b
    + \tfrac16 \Theta_{abc}\eta_a\eta_b\eta_c + O(\delta^4),
    \label{eq:C1}
\end{equation}
where the gradient $g_a$, Hessian $H_{ab}$ and third-derivative tensor
$\Theta_{abc}$ are evaluated at $\bm J = J_0\mathbf{1}$ and repeated bond indices are summed. We write $J^*_\omega \equiv J^*(\bm J^\omega)$ for the value at
realization $\omega$.

The linear coefficient is fixed as follows. A teacher whose bonds are all shifted by the same $\epsilon$ remains inside the student's uniform hypothesis class, so it is recovered exactly as
\begin{equation}
    J^*\big((J_0+\epsilon)\mathbf{1}\big) = J_0 + \epsilon ,
    \label{eq:C2}
\end{equation}

Taking the derivative of Eq.~\eqref{eq:C2} with respect to $\epsilon$ gives

\begin{equation}
\sum_{\alpha=1}^{N_b}
\left.\frac{\partial J^*}{\partial J_\alpha}\right|_{\bm J = J_0\mathbf{1}}
\frac{dJ_\alpha}{d\epsilon}
= \sum_{\alpha=1}^{N_b} g_\alpha \cdot 1
= \sum_{\alpha=1}^{N_b} g_\alpha = 1 .
\label{eq:C3}
\end{equation}

The clean lattice is homogeneous, so all bonds are equivalent and the $g_a$ are equal. Together,
\begin{equation}
    g_a = \left.\frac{\partial J^*}{\partial J_a}\right|_{\bm J = J_0\mathbf{1}}
    = \frac{1}{N_b},
        \label{eq:C4}
\end{equation}

\subsection{Bias of the system}
\label{sec:appendix_bias}

In the infinite data limit, bias associated with a given teacher $\omega$ is defined as $b = \mathbb{E}\!\left[J^*(\boldsymbol{J)}\right] - J_0$. Here, $\langle \eta \rangle = 0$ and the dominating term is coming from the Hessian following $\langle \eta^2 \rangle = \frac{\delta^2}{3}$.

\begin{align}
b
&= \frac{\delta^2}{6} H_{ab}\delta_{ab}
   + O(\delta^4) \\
&= \frac{\delta^2}{6}\operatorname{Tr} H(T)
   + O(\delta^4) \\
&\approx \frac{N_b\delta^2}{6} H_{aa}.
\label{eq:C7}
\end{align}

\subsection{Disorder Variance}
\label{sec:disorder_variance_appendix}

The expansion in Eq.~\eqref{eq:C1} is for a given teacher $\omega$. But the disorder variance is averaged over all the teachers. So, it is necessary to write Eq.~\eqref{eq:C1} as

\begin{equation}
    J^*_\omega(\boldsymbol{J}) - J_0 = L + Q + C + O(\delta^4),
    \label{eq:C8}
\end{equation}
with
\begin{equation}
    L = g_a\eta_a,
    \qquad
    Q = \tfrac{1}{2}H_{ab}\,\eta_a\eta_b,
    \qquad
    C = \tfrac{1}{6}\Theta_{abc}\,\eta_a\eta_b\eta_c ,
    \label{eq:C9}
\end{equation}
summation over repeated bond indices implied. Expanding the variance of Eq.~\eqref{eq:C8} in full,
\begin{align}
    V_{\mathrm{disorder}}
    = \operatorname{Var}(L) &+ \operatorname{Var}(Q) + \operatorname{Var}(C) \nonumber\\
    &+ 2\operatorname{Cov}(L,Q) + 2\operatorname{Cov}(L,C)
     + 2\operatorname{Cov}(Q,C),
    \label{eq:C10}
\end{align}
whose terms are of order $\delta^2$, $\delta^4$, $\delta^6$, $\delta^3$,
$\delta^4$ and $\delta^5$. Since $\eta$ are taken from uniform distribution, only even order terms survive, giving

\begin{equation}
    V_{\mathrm{disorder}}
    = \operatorname{Var}(L) + \operatorname{Var}(Q)
    + 2\operatorname{Cov}(L,C) + O(\delta^6).
    \label{eq:C11}
\end{equation}

For $\eta_a\sim\mathcal{U}(-\delta,\delta)$,
\begin{equation}
    \mu_2 = \frac{1}{2\delta}\int_{-\delta}^{\delta}\!\eta^2\,d\eta
    = \frac{\delta^2}{3},
    \qquad
    \mu_4 = \frac{1}{2\delta}\int_{-\delta}^{\delta}\!\eta^4\,d\eta
    = \frac{\delta^4}{5},
    \label{eq:C12}
\end{equation}
with kurtosis
\begin{equation}
    \kappa_4 \equiv \mu_4 - 3\mu_2^2
    = \frac{\delta^4}{5} - \frac{\delta^4}{3}
    = -\frac{2\delta^4}{15}.
    \label{eq:C13}
\end{equation}
For four bond indices, since all $\eta$ are independent implies that the
average is nonzero only when the indices pair up, giving
\begin{equation}
    \langle \eta_a\eta_b\eta_c\eta_d\rangle
    = \mu_2^2\big(\delta_{ab}\delta_{cd}+\delta_{ac}\delta_{bd}
    +\delta_{ad}\delta_{bc}\big) + \kappa_4\,\delta_{abcd},
    \label{eq:C14}
\end{equation}
where $\delta_{abcd}=1$ only for $a=b=c=d$. The three Kronecker
products are the Wick pairings appropriate to a Gaussian and the $\kappa_4$ term is the non-Gaussian correction.

With $g_a = 1/N_b$ from Eq.~\eqref{eq:C4},
\begin{equation}
    \operatorname{Var}(L) = g_ag_b\langle\eta_a\eta_b\rangle
    = \mu_2\sum_a g_a^2
    = \mu_2 N_b \frac{1}{N_b^2}
    = \frac{\delta^2}{3N_b},
    \label{eq:C15}
\end{equation}

For the variance of the Hessian , 

\begin{equation}
    \operatorname{Var}(Q)
    = \langle Q^2\rangle - \langle Q\rangle^2 .
    \label{eq:C16}
\end{equation}

The first moment is
\begin{equation}
    \langle Q\rangle
    = \frac{1}{2}H_{ab}\,\mu_2\delta_{ab}
    = \frac{\mu_2}{2}\operatorname{Tr}H .
    \label{eq:C17}
\end{equation}

The second moment is
\begin{align}
    \langle Q^2\rangle
    &= \frac{1}{4}H_{ab}H_{cd}
       \langle\eta_a\eta_b\eta_c\eta_d\rangle \\
    &= \frac{\mu_2^2}{4}
       \left[
       \underbrace{H_{aa}H_{cc}}_{(\operatorname{Tr}H)^2}
       + \underbrace{H_{ab}H_{ab}}_{\operatorname{Tr}(H^2)}
       + \underbrace{H_{ab}H_{ba}}_{\operatorname{Tr}(H^2)}
       \right]
       + \frac{\kappa_4}{4}\sum_a H_{aa}^2 .
    \label{eq:C19}
\end{align}

Subtracting $\langle Q\rangle^2$ from $\langle Q^2\rangle$ gives
\begin{equation}
    \operatorname{Var}(Q)
    = \frac{\mu_2^2}{2}\operatorname{Tr}(H^2)
    + \frac{\kappa_4}{4}\sum_a H_{aa}^2 .
    \label{eq:C20}
\end{equation}

For $\mu_2=\delta^2/3$ and $\kappa_4=-2\delta^4/15$,
this becomes
\begin{equation}
    \operatorname{Var}(Q)
    = \delta^4
    \left[
        \frac{1}{18}\operatorname{Tr}(H^2)
        - \frac{1}{30}\sum_a H_{aa}^2
    \right]
    \equiv \delta^4 C_4^{(H)} .
    \label{eq:C21}
\end{equation}
which is Eq.~\eqref{eq:52}.

Now, for the covariance between the gradient and third derivative, since $\langle L\rangle = \langle C\rangle=0$, one can write $\operatorname{Cov}(L,C) = \langle LC\rangle$ 
\begin{equation}
    \langle LC\rangle
    = \frac{1}{6}\,g_d\,\Theta_{abc}\,\langle\eta_d\eta_a\eta_b\eta_c\rangle .
    \label{eq:C22}
\end{equation}
Because $\Theta_{abc}$ is fully symmetric, the three pairings contract to the same object:
\begin{equation}
    g_d\Theta_{abc}\big(\delta_{da}\delta_{bc}+\delta_{db}\delta_{ac}
    +\delta_{dc}\delta_{ab}\big) = 3\sum_{a,c}g_a\Theta_{acc},
    \label{eq:C23}
\end{equation}
while the kurtosis term contributes $\sum_a g_a\Theta_{aaa}$. Hence
\begin{equation}
    \operatorname{Cov}(L,C)
    = \frac{1}{6}\Big[3\mu_2^2\sum_{a,c}g_a\Theta_{acc}
    + \kappa_4\sum_a g_a\Theta_{aaa}\Big].
    \label{eq:C24}
\end{equation}
Using $g_a = 1/N_b$ and substituting $\mu_2^2=\delta^4/9$
and $\kappa_4/3 = -2\delta^4/45$,
\begin{equation}
    2\operatorname{Cov}(L,C) = \delta^4\left[
    \frac{1}{9N_b}\sum_{a,c}\Theta_{acc}
    - \frac{2}{45N_b}\sum_a \Theta_{aaa}\right]
    \equiv \delta^4 C_4^{(3)},
    \label{eq:C25}
\end{equation}

\subsection{The scaling form}
\label{sec:scaling_form_appendix}

The Boltzmann weight can be written entirely in terms of the reduced couplings $K_a = \beta J_a$,
\begin{equation}
    P(\boldsymbol{\sigma})
    = \frac{1}{Z}
      \exp\left(\sum_a K_a X_a\right),
    \qquad
    X_a = \sigma_i\sigma_j
    \label{eq:C26}
\end{equation}
so temperature and couplings never appear separately. Thus every equilibrium average is therefore a function of $\bm K$ including the pattern frequencies $p(x,\boldsymbol{K})$.

For a given teacher $\omega$, the estimator $J^*$ is given by

\begin{equation}
    J^* = \frac{1}{\beta}\,\varphi,
    \qquad
    \varphi = \frac{\sum_x p(x)S(x)f(x)}{\sum_x p(x)S(x)^2}.
    \label{eq:C27}
\end{equation}
No $\beta$ enters in $\varphi$ itself. But the pattern frequency $p$ depends on temperature such that $p(x) = p = p(\boldsymbol{K})$. Thus defining the composition $\Phi(\bm K) \equiv \varphi[p(\bm K)]$, which is dimensionless, one can write 

\begin{equation}
    J^*(J,T) = T\,\Phi(\beta\bm J)
    \label{eq:C28}
\end{equation}

Since $K_b = \beta J_b$ we have $\partial/\partial J_a = \beta\,\partial/\partial
K_a$. Thus
\begin{align}
    g_a &= \frac{1}{\beta}\cdot\beta\,\partial_{K_a}\Phi
         = \Phi'_a(\bm K),
    \label{eq:C29}\\
    H_{ab} &= \frac{1}{\beta}\cdot\beta^2\,\partial_{K_a}\partial_{K_b}\Phi
         = \beta\,\Phi''_{ab}(\bm K),
    \label{eq:C30}\\
    \Theta_{abc} &= \frac{1}{\beta}\cdot\beta^3\,
    \partial_{K_a}\partial_{K_b}\partial_{K_c}\Phi
         = \beta^2\,\Phi'''_{abc}(\bm K),
    \label{eq:C31}
\end{align}
and in general $\partial^n_{\bm J}J^* = \beta^{\,n-1}\Phi^{(n)}(\bm K)$. This temperature dependence of the terms shows that in the limit $T  \rightarrow \infty$, only the $g_a\eta_a$ term survives. And the disorder variance hits the floor $\frac{\delta^2}{3 N_b}$.

\section{Fixed-pattern approximation to the bias}
\label{app:bias_derivation}

We derive the equation for bias in the fixed pattern condition, which means when the student gets a finite number of configurations.

Consider a site in the 2D Ising model whose coordination number $z = 4$, nearest neighbours form the pattern $x = \{\sigma_1, \sigma_2,\sigma_3,\sigma_4\}$, with neighbour sum $S(x) = \sum_k \sigma_k$. The conditional probability
that the central spin is $+1$ given its neighbours is
\begin{equation}
    P(\sigma = +1 \mid \{\sigma_k\}) = \Phi\!\left(2\beta \sum_{k=1}^{z} J_k \sigma_k\right),
    \label{eq:D1}
\end{equation}
where $\Phi(u) = 1/(1+e^{-u})$ is the sigmoid function, $\beta = 1/T$, and the bonds $J_k = J_0 + \eta_k$ are drawn independently with $\eta_k \sim \mathcal{U}(-\delta, +\delta)$. The argument of the sigmoid splits into a deterministic and a random part,
\begin{equation}
    \begin{aligned}
        2\beta \sum_k J_k \sigma_k
        &= 2\beta J_0 S(x) + 2\beta \sum_k \eta_k \sigma_k \\[4pt]
        &= \bar\xi + \Delta\xi,
    \end{aligned}
    \label{eq:D2}
\end{equation}
where $\bar\xi$ is fixed by the pattern and $\Delta\xi$ is a zero-mean random
variable capturing the bond disorder.

The student observes empirical pattern statistics, from which it
infers a local field
\begin{equation}
    f(x) = \tfrac{1}{2}\ln \frac{{P}(+1 \mid x)}{1 - {P}(+1 \mid x)}.
    \label{eq:D3}
\end{equation}
Within this local-bond expansion we average the incident bond perturbations at fixed pattern, giving
$\bar{P}(+1 \mid x) \simeq \mathbb{E}_\eta[\Phi(\bar\xi + \Delta\xi)]$.
The approximation neglects the disorder dependence of the probability with which the pattern itself occurs.

Taylor expanding $\Phi$ about $\bar\xi$,
\begin{equation}
    \Phi(\bar\xi + \Delta\xi) = \Phi(\bar\xi)
    + \Phi'(\bar\xi)\,\Delta\xi
    + \tfrac{1}{2}\Phi''(\bar\xi)(\Delta\xi)^2 + O((\Delta\xi)^3),
    \label{eq:D4}
\end{equation}
and taking the disorder average,
\begin{equation}
    \bar{P}(+1 \mid x) = \Phi(\bar\xi)
    + \tfrac{1}{2}\Phi''(\bar\xi)\,\langle(\Delta\xi)^2\rangle
    + O(\delta^4).
    \label{eq:D5}
\end{equation}
Only even moments of $\Delta\xi$ contribute, since odd moments vanish
under the symmetry of $\eta$. The second moment is
\begin{equation}
    \langle(\Delta\xi)^2\rangle
    = 4\beta^2 \sum_k \sigma_k^2 \,\langle\eta_k^2\rangle
    = \frac{4 z \beta^2 \delta^2}{3}
    = \frac{16 \beta^2 \delta^2}{3},
    \label{eq:D6}
\end{equation}
using $\sigma_k^2 = 1$, $\langle\eta_k^2\rangle = \delta^2/3$ for the
uniform distribution, and $z = 4$.

Substituting Eq.~\eqref{eq:D6} into
Eq.~\eqref{eq:D5} gives the leading shift in $\bar P$,
\begin{equation}
    \bar P(+1 \mid x) = \Phi(\bar\xi)
    + \frac{8 \beta^2 \delta^2}{3}\,\Phi''(\bar\xi)
    + O(\delta^4).
    \label{eq:D7}
\end{equation}

To simplify $\Phi''(\bar\xi)$ we use two standard identities for
the logistic sigmoid,
\begin{align}
    \Phi'(u) &= \Phi(u)[1 - \Phi(u)], \\
    1 - 2\Phi(u) &= -\tanh(u/2).
    \label{eq:D9}
\end{align}
Differentiating $\Phi'(u)$ and using the second identity,
\begin{equation}
    \Phi''(u) = \Phi'(u)[1 - 2\Phi(u)] = -\Phi'(u)\,\tanh(u/2),
    \label{eq:D10}
\end{equation}
so that at $u = \bar\xi = 2\beta J_0 S(x)$,
\begin{equation}
    \Phi''(\bar\xi) = -\Phi'(\bar\xi)\,\tanh(\beta J_0 S(x)).
    \label{eq:D11}
\end{equation}
Inserting Eq.~\eqref{eq:D11} into
Eq.~\eqref{eq:D7} yields the disorder-induced shift in
$\bar P$,
\begin{equation}
    \begin{aligned}
        \bar P(+1 \mid x)
        &= \Phi(\bar\xi)
        - \frac{8\beta^2\delta^2}{3}\,\Phi'(\bar\xi)\,
          \tanh\!\big(\beta J_0 S(x)\big)
        + O(\delta^4) \\[4pt]
        &\equiv \Phi(\bar\xi) + \epsilon .
    \end{aligned}
    \label{eq:D12}
\end{equation}

The disorder-averaged field $\bar{f}(x)$ is then
\begin{equation}
    \bar f(x) = \tfrac{1}{2}\ln\frac{\Phi(\bar\xi) + \epsilon}{1 - \Phi(\bar\xi) - \epsilon}.
    \label{eq:D13}
\end{equation}
Splitting the logarithm and expanding $\ln(1 + u) = u + O(u^2)$ to
first order in $\epsilon$,

\begin{equation}
    \begin{aligned}
        \bar f(x)
        &= \tfrac{1}{2}\ln\frac{\Phi(\bar\xi)}{1 - \Phi(\bar\xi)}
           + \frac{\epsilon}{2}\left[\frac{1}{\Phi(\bar\xi)}
             + \frac{1}{1 - \Phi(\bar\xi)}\right]
           + O(\epsilon^2) \\[6pt]
        &= \beta J_0 S(x)
           + \frac{\epsilon}{2\,\Phi'(\bar\xi)} + O(\delta^4) \\[6pt]
        &= \beta J_0 S(x)
           - \frac{4\beta^2 \delta^2}{3}\,\tanh(\beta J_0 S(x))
           + O(\delta^4).
    \end{aligned}
    \label{eq:D14}
\end{equation}
The first term is the clean-system field and the second is the
disorder-induced correction. The correction is proportional to
$\delta^2$ and vanishes at $S(x) = 0$.

The inferred coupling is obtained by projection,
\begin{equation}
    \hat{J} = \frac{\sum_x N(x)\,S(x)\,\bar{f}(x)}{\beta\sum_x N(x)\,S(x)^2},
    \label{eq:D15}
\end{equation}
and substituting Eq.~\eqref{eq:D14} gives
\begin{equation}
    \mathbb{E}[\hat{J}] = J_0
    - \frac{4\delta^2}{3T}\,
    \frac{\sum_x N(x)\,S(x)\tanh(\beta J_0 S(x))}{\sum_x N(x)\,S(x)^2}
    + O(\delta^4).
    \label{eq:D16}
\end{equation}
Identifying
\begin{equation}
    K_{\rm loc}(T) = \frac{4}{3T}\,
    \frac{\sum_x N(x)\,S(x)\tanh(\beta J_0 S(x))}{\sum_x N(x)\,S(x)^2}
    \label{eq:D17}
\end{equation}
gives the leading local-bond estimate of the bias,
\begin{equation}
    b_{\rm loc}(\delta,T)
    =-K_{\rm loc}(T)\,\delta^2+O(\delta^4).
    \label{eq:D18}
\end{equation}

The same fixed-pattern approximation gives an estimate of the derivative of the local contribution with respect to the mean coupling \(J_0\). Differentiating
Eq.~\eqref{eq:D18} at fixed \(T\) and \(\delta\), and using
$\partial\tanh(u)/\partial u = \mathrm{sech}^2(u)$ with
$u = \beta J_0 S(x)$, gives
\begin{equation}
    b_{\rm loc}'(J_0)
    = -\frac{z\,\delta^2}{3T^2}\,
    \frac{\sum_x N(x)\,S(x)^2\,\mathrm{sech}^2(\beta J_0 S(x))}{\sum_x N(x)\,S(x)^2},
    \label{eq:D19}
\end{equation}
with \(z=4\). 

\section{Kernel framework and residual complexity}
\label{app:kernels_RC}

This appendix describes the kernel regression framework used to
infer the local field, defines the residual complexity that
quantifies the inadequacy of the uniform-coupling description.

Consider a student $s$, inferring a coupling $\hat{J}_{\omega,s}$ from $M$ configurations from a given teacher realization $\omega$. Firstly, in order to obtain the empirical local fields $f(x)$, we fit a regularized model of the form
\begin{equation}
    \tilde{f}(x) = \sum_{x'=0}^{15} \alpha_{x'}\, K(x, x'),
    \label{eq:E1}
\end{equation}
where $K(x, x')$ is a kernel function measuring the similarity
between patterns $x$ and $x'$, and
$\boldsymbol{\alpha} = (\alpha_0, \ldots, \alpha_{15})^\top$ is the
vector of dual coefficients~\cite{Schlkopf2001}. In all four cases 16 coefficients are fitted, and the rank of the kernel matrix $\mathbf{K}$ constrains the dimensionality of the resulting function. The
coefficients are determined by minimizing the regularized weighted
squared error,
\begin{equation}
    \mathcal{L}(\boldsymbol{\alpha}) = \sum_{x=0}^{15} N(x)\,\big[\tilde{f}(x) - f(x)\big]^2
    + \lambda\,\boldsymbol{\alpha}^\top \mathbf{K}\,\boldsymbol{\alpha},
    \label{eq:E2}
\end{equation}
with closed-form solution ~\cite{Schlkopf2001,Rasmussen2004}
\begin{equation}
    \boldsymbol{\alpha}^* = (\mathbf{W}\mathbf{K} + \lambda \mathbf{I})^{-1}\,\mathbf{W}\,\mathbf{f},
    \label{eq:E3}
\end{equation}
where $\mathbf{W} = \mathrm{diag}(N(0), \ldots, N(15))$ and
$\mathbf{f} = (f(0),\ldots,f(15))^\top$. The fitted field is $\tilde{\mathbf{f}} = \mathbf{K}\,\boldsymbol{\alpha}^*$,
from which $\hat{J}_{\omega,s}$ is computed.

Each kernel encodes a different prior assumption about the structure of
the local field. The \textit{identity} kernel,
$K_{\mathrm{id}}(x,x') = \delta_{xx'}$, makes no assumption at all such that each pattern is treated independently with $\tilde{f}(x) = f(x)$, so every deviation in the local field is retained. This is the most flexible
choice. The \textit{linear} kernel,
$K_{\mathrm{lin}}(x,x') = \boldsymbol{\phi}(x)\cdot\boldsymbol{\phi}(x')$
with $\boldsymbol{\phi}(x) = (\sigma_1(x), \ldots, \sigma_4(x))^\top$ the
spin vector of pattern $x$, has rank $4$ and restricts the field to
$\tilde{f}(x) = \sum_j w_j \sigma_j(x)$, representing direction-dependent
effective couplings. This is the space of functions realized by anisotropic bonds in an Ising model. The \textit{isotropic} kernel,
$K_{\mathrm{iso}}(x,x') = S(x)S(x')$, has rank $1$ and constrains the
field to the uniform-coupling form $\tilde f(x) = \beta J\,S(x)$, for which the
residual complexity is zero by construction. Finally, the
\textit{radial basis function} (RBF) kernel,
$K_{\mathrm{RBF}}(x,x') = \exp\!\big(-|\boldsymbol{\phi}(x) -
\boldsymbol{\phi}(x')|^2 / 2\ell^2\big)$, is full rank for finite \(\ell\). In the small-\(\ell\) limit it approaches an identity-like kernel, whereas the large-\(\ell\) limit, it becomes equal to the linear kernel. We therefore use the RBF kernel only as a flexible full-rank cross-check of the projection.

\begin{figure*}[t!]
   
    \includegraphics[width=0.99\linewidth]{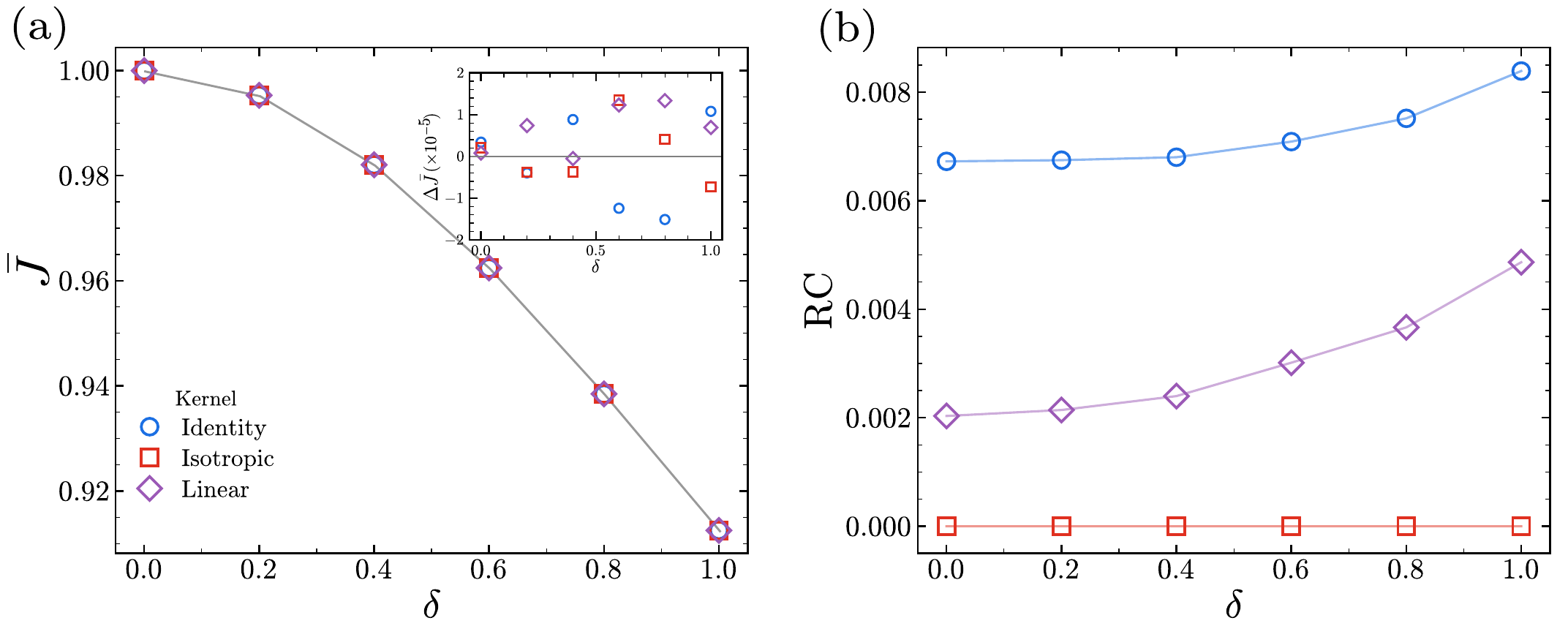}
    \caption{Inferred coupling and residual complexity across kernels.
    \textbf{(a)} Inferred coupling $\bar{J}$, averaged across all students and teachers, at $T \approx 2.20$ as a function of disorder strength $\delta$, computed with three kernels:
    identity, isotropic, and linear. All three kernels yield
    the same $\bar{J}$ within numerical precision, as demonstrated by
    the inset showing the deviation $\Delta\bar{J}$ of each kernel from the mean of all three. The spread across kernels is of order $10^{-5}$.
    \textbf{(b)} Residual complexity $\mathrm{RC}$ at $T \approx 2.20$
    as a function of $\delta$ for the same three kernels. The isotropic
    kernel yields $\mathrm{RC} = 0$ identically, since it retains only
    the component of the field aligned with $S(x)$. The linear kernel
    gives a small but growing $\mathrm{RC}$ that captures the
    directional anisotropy of the empirical field. The identity kernel yield the largest $\mathrm{RC}$, which additionally captures the disorder-induced nonlinear corrections. All RCs grow with disorder, reflecting the growing mismatch between the uniform-coupling model and the heterogeneous teacher.}
    \label{fig:6}
\end{figure*}

Figure~\ref{fig:6}(a) shows the inferred coupling $\bar J$, averaged over
all students and teachers, obtained with each of the four kernels. It is
essentially independent of the kernel choice, with a maximum spread of order
$10^{-5}$, and this follows from the structure of the projection. Viewing a field
on the sixteen patterns as a vector in $\mathbb{R}^{16}$ with the inner product
$\langle u,v\rangle_N = \sum_x N(x)u(x)v(x)$, which weights each pattern by how
often it is observed, the estimator of Eq.~\eqref{eq:16} reads
\begin{equation}
    \hat{J}_{\omega,s}
    = \frac{\langle \tilde f, S\rangle_N}{\beta\,\langle S, S\rangle_N},
    \label{eq:E4}
\end{equation}
so the coupling depends on all sixteen components of the fitted field through a single scalar. At $\lambda\to0$ $\tilde{f}$ is the best weighted fit to $f$ among the fields the kernel can represent and the residual $f - \tilde f$ is orthogonal to every field the kernel can represent leading to 

\begin{equation}
    \langle \tilde f, S\rangle_N = \langle f, S\rangle_N ,
    \label{eq:E5}
\end{equation}
and by Eq.~\eqref{eq:E4} all four return the same coupling,
\begin{equation}
    \hat{J}_{\mathrm{id}} = \hat{J}_{\mathrm{lin}}
    = \hat{J}_{\mathrm{iso}}
    \label{eq:kernel_equality}
\end{equation}

Writing the disorder-averaged empirical field as its clean-Ising part plus a remainder,
\begin{equation}
\begin{aligned}
    f(x) &= \beta J_0 S(x) + \rho(x),\\
    \rho(x) &= -\tfrac{4}{3}\beta^2\delta^2
    \tanh\!\big(\beta J_0S(x)\big) + O(\delta^4).
\end{aligned}
\label{eq:E7}
\end{equation}

with $\rho$ from Eq.~\eqref{eq:D14}, and splitting $\rho$ into its components
parallel and perpendicular to $S$,
\begin{equation}
    f = \beta J_0 S + \beta b\, S + r,
    \qquad \langle r, S\rangle_N = 0,
    \label{eq:E8}
\end{equation}
substitution into Eq.~\eqref{eq:E4} gives
\begin{equation}
    \hat{J}_{\omega,s} = J_0 + b + 0,
    \qquad
    b = \frac{\langle \rho, S\rangle_N}{\beta\,\langle S,S\rangle_N},
    \label{eq:E9}
\end{equation}
where $b$ denotes the bias. The bias is thus exactly the component of $\rho$ along $S$. Since $S\tanh(\beta J_0S) > 0$ for every $S\neq0$ makes every term of $\langle\rho,S\rangle_N$ negative. The third term vanishes because $r$ is orthogonal to $S$ by construction, so the residual however large cannot affect the inferred coupling. The residual can be written as
\begin{equation}
    r(x) = \tilde{f}(x) - \beta\hat{J}_{\omega,s}\, S(x),
    \label{eq:E10}
\end{equation}

We summarize the discarded part by a single scalar, the residual
complexity,
\begin{equation}
    \mathrm{RC} = \sqrt{\frac{\sum_x N(x)\,
    \bigl[\tilde{f}(x) - \beta\hat{J}_{\omega,s} S(x)\bigr]^2}{\sum_x N(x)}},
    \label{eq:E11}
\end{equation}

$\mathrm{RC}$ measures the structure lost during the projection. The kernel choice controls what portion of the field enters the RC, and hence which physical mechanisms the RC measures. The dependence of the RC on disorder strength $\delta$ at $T=2.20$ is shown in Fig.~\ref{fig:6}b for the different kernels. The identity kernel gives the largest RC, since it assumes no functional form for $\tilde{f}(x)$ and captures all interactions. But since we project this $f(x)$ to a single effective coupling, it discards the most information, followed by the linear kernel and then by the isotropic kernel. For isotropic kernel, $f(x)$ is evaluated by the assumption that all the bonds are identical, so projection to clean Ising model results in $\mathrm{RC} = 0$. Like bias, RC can be driven to zero only by enlarging the hypothesis class, ie, projecting to the full disordered Ising space instead of the clean Ising subspace.

\bibliography{apssamp}

@article{Chayes1984,
  author  = {Chayes, J. T. and Chayes, L. and Lieb, Elliott H.},
  title   = {The inverse problem in classical statistical mechanics},
  journal = {Commun. Math. Phys.},
  volume  = {93},
  pages   = {57},
  year    = {1984},
  doi     = {10.1007/BF01218639}
}

@article{Nguyen_2017,
  author  = {Nguyen, H. Chau and Zecchina, Riccardo and Berg, Johannes},
  title   = {Inverse statistical problems: from the inverse {Ising} problem to data science},
  journal = {Adv. Phys.},
  volume  = {66},
  pages   = {197},
  year    = {2017},
  doi     = {10.1080/00018732.2017.1341604}
}

@article{Kunkin1969,
  author  = {Kunkin, W. and Frisch, H. L.},
  title   = {Inverse problem in classical statistical mechanics},
  journal = {Phys. Rev.},
  volume  = {177},
  pages   = {282},
  year    = {1969},
  doi     = {10.1103/PhysRev.177.282}
}

@article{Changlani2015,
  author  = {Changlani, Hitesh J. and Zheng, Huihuo and Wagner, Lucas K.},
  title   = {Density-matrix based determination of low-energy model {Hamiltonians} from \textit{ab initio} wavefunctions},
  journal = {J. Chem. Phys.},
  volume  = {143},
  pages   = {102814},
  year    = {2015},
  doi     = {10.1063/1.4927664}
}

@article{Krumsiek2011,
  author  = {Krumsiek, Jan and Suhre, Karsten and Illig, Thomas and Adamski, Jerzy and Theis, Fabian J.},
  title   = {{Gaussian} graphical modeling reconstructs pathway reactions from high-throughput metabolomics data},
  journal = {BMC Syst. Biol.},
  volume  = {5},
  pages   = {21},
  year    = {2011},
  doi     = {10.1186/1752-0509-5-21}
}

@article{Cocco2009,
  author  = {Cocco, Simona and Leibler, Stanislas and Monasson, R{\'e}mi},
  title   = {Neuronal couplings between retinal ganglion cells inferred by efficient inverse statistical physics methods},
  journal = {Proc. Natl. Acad. Sci. U.S.A.},
  volume  = {106},
  pages   = {14058},
  year    = {2009},
  doi     = {10.1073/pnas.0906705106}
}

@article{Roudi2009,
  author  = {Roudi, Yasser and Nirenberg, Sheila and Latham, Peter E.},
  title   = {Pairwise maximum entropy models for studying large biological systems: When they can work and when they can't},
  journal = {PLoS Comput. Biol.},
  volume  = {5},
  pages   = {e1000380},
  year    = {2009},
  doi     = {10.1371/journal.pcbi.1000380}
}

@article{Wang2017,
  author  = {Wang, Sheng and Sun, Siqi and Li, Zhen and Zhang, Renyu and Xu, Jinbo},
  title   = {Accurate \textit{de novo} prediction of protein contact map by ultra-deep learning model},
  journal = {PLoS Comput. Biol.},
  volume  = {13},
  pages   = {e1005324},
  year    = {2017},
  doi     = {10.1371/journal.pcbi.1005324}
}

@article{Lezon2006,
  author  = {Lezon, Timothy R. and Banavar, Jayanth R. and Cieplak, Marek and Maritan, Amos and Fedoroff, Nina V.},
  title   = {Using the principle of entropy maximization to infer genetic interaction networks from gene expression patterns},
  journal = {Proc. Natl. Acad. Sci. U.S.A.},
  volume  = {103},
  pages   = {19033},
  year    = {2006},
  doi     = {10.1073/pnas.0609152103}
}

@article{Bialek,
  author  = {Bialek, William and Cavagna, Andrea and Giardina, Irene and Mora, Thierry and Pohl, Oliver and Silvestri, Edmondo and Viale, Massimiliano and Walczak, Aleksandra M.},
  title   = {Social interactions dominate speed control in poising natural flocks near criticality},
  journal = {Proc. Natl. Acad. Sci. U.S.A.},
  volume  = {111},
  pages   = {7212},
  year    = {2014},
  doi     = {10.1073/pnas.1324045111}
}

@inproceedings{Kappen,
  author    = {Kappen, Hilbert J. and Rodr{\'\i}guez, Francisco B.},
  title     = {{Boltzmann} machine learning using mean field theory and linear response correction},
  booktitle = {Advances in Neural Information Processing Systems 10},
  editor    = {Jordan, M. I. and Kearns, M. J. and Solla, S. A.},
  publisher = {MIT Press},
  address   = {Cambridge, MA},
  pages     = {280},
  year      = {1998}
}

@article{NguyenBerg2012,
  author  = {Nguyen, H. Chau and Berg, Johannes},
  title   = {Mean-field theory for the inverse {Ising} problem at low temperatures},
  journal = {Phys. Rev. Lett.},
  volume  = {109},
  pages   = {050602},
  year    = {2012},
  doi     = {10.1103/PhysRevLett.109.050602}
}

@article{Cocco_2011,
  author  = {Cocco, S. and Monasson, R.},
  title   = {Adaptive cluster expansion for inferring {Boltzmann} machines with noisy data},
  journal = {Phys. Rev. Lett.},
  volume  = {106},
  pages   = {090601},
  year    = {2011},
  doi     = {10.1103/PhysRevLett.106.090601}
}

@article{Aurell2012,
  author  = {Aurell, Erik and Ekeberg, Magnus},
  title   = {Inverse {Ising} inference using all the data},
  journal = {Phys. Rev. Lett.},
  volume  = {108},
  pages   = {090201},
  year    = {2012},
  doi     = {10.1103/PhysRevLett.108.090201}
}

@article{Hyvrinen2006ConsistencyOP,
  author  = {Hyv{\"a}rinen, Aapo},
  title   = {Consistency of pseudolikelihood estimation of fully visible {Boltzmann} machines},
  journal = {Neural Comput.},
  volume  = {18},
  pages   = {2283},
  year    = {2006},
  doi     = {10.1162/neco.2006.18.10.2283}
}

@misc{Lakshminarayanan2017,
  author        = {Lakshminarayanan, Balaji and Pritzel, Alexander and Blundell, Charles},
  title         = {Simple and scalable predictive uncertainty estimation using deep ensembles},
  year          = {2017},
  eprint        = {1612.01474},
  archivePrefix = {arXiv},
  primaryClass  = {stat.ML}
}

@misc{Fort2019,
  author        = {Fort, Stanislav and Hu, Huiyi and Lakshminarayanan, Balaji},
  title         = {Deep ensembles: A loss landscape perspective},
  year          = {2019},
  eprint        = {1912.02757},
  archivePrefix = {arXiv},
  primaryClass  = {stat.ML}
}

@article{Ngampruetikorn_2022,
  author  = {Ngampruetikorn, Vudtiwat and Sachdeva, Vedant and Torrence, Johanna and Humplik, Jan and Schwab, David J. and Palmer, Stephanie E.},
  title   = {Inferring couplings in networks across order-disorder phase transitions},
  journal = {Phys. Rev. Res.},
  volume  = {4},
  pages   = {023240},
  year    = {2022},
  doi     = {10.1103/PhysRevResearch.4.023240}
}

@article{Loureiro_2022,
  author  = {Loureiro, Bruno and Gerbelot, C{\'e}dric and Cui, Hugo and Goldt, Sebastian and Krzakala, Florent and M{\'e}zard, Marc and Zdeborov{\'a}, Lenka},
  title   = {Learning curves of generic features maps for realistic datasets with a teacher-student model},
  journal = {J. Stat. Mech.},
  volume  = {2022},
  pages   = {114001},
  year    = {2022},
  doi     = {10.1088/1742-5468/ac9825}
}

@article{Besag1974,
  author  = {Besag, Julian},
  title   = {Spatial interaction and the statistical analysis of lattice systems},
  journal = {J. R. Stat. Soc. Ser. B},
  volume  = {36},
  pages   = {192},
  year    = {1974},
  doi     = {10.1111/j.2517-6161.1974.tb00999.x}
}

@article{White1982,
  author  = {White, Halbert},
  title   = {Maximum likelihood estimation of misspecified models},
  journal = {Econometrica},
  volume  = {50},
  pages   = {1},
  year    = {1982},
  doi     = {10.2307/1912526}
}

@article{Jeffreys1946,
  author  = {Jeffreys, Harold},
  title   = {An invariant form for the prior probability in estimation problems},
  journal = {Proc. R. Soc. London, Ser. A},
  volume  = {186},
  pages   = {453},
  year    = {1946},
  doi     = {10.1098/rspa.1946.0056}
}

@article{Harris1974,
  author  = {Harris, A. B.},
  title   = {Effect of random defects on the critical behaviour of {Ising} models},
  journal = {J. Phys. C},
  volume  = {7},
  pages   = {1671},
  year    = {1974},
  doi     = {10.1088/0022-3719/7/9/009}
}

@book{MacKay2004InformationTI,
  author    = {MacKay, David J. C.},
  title     = {Information Theory, Inference, and Learning Algorithms},
  publisher = {Cambridge University Press},
  address   = {Cambridge},
  year      = {2003}
}

@article{CR_bound,
  author  = {Kullback, S.},
  title   = {Certain inequalities in information theory and the {Cram{\'e}r--Rao} inequality},
  journal = {Ann. Math. Stat.},
  volume  = {25},
  pages   = {745},
  year    = {1954},
  doi     = {10.1214/aoms/1177728660}
}

@article{Aharony1996,
  author  = {Aharony, Amnon and Harris, A. Brooks},
  title   = {Absence of self-averaging and universal fluctuations in random systems near critical points},
  journal = {Phys. Rev. Lett.},
  volume  = {77},
  pages   = {3700},
  year    = {1996},
  doi     = {10.1103/PhysRevLett.77.3700}
}

@article{Machta2013,
  author  = {Machta, Benjamin B. and Chachra, Ricky and Transtrum, Mark K. and Sethna, James P.},
  title   = {Parameter space compression underlies emergent theories and predictive models},
  journal = {Science},
  volume  = {342},
  pages   = {604},
  year    = {2013},
  doi     = {10.1126/science.1238723}
}

@article{Wolff,
  author  = {Wolff, Ulli},
  title   = {Collective {Monte Carlo} updating for spin systems},
  journal = {Phys. Rev. Lett.},
  volume  = {62},
  pages   = {361},
  year    = {1989},
  doi     = {10.1103/PhysRevLett.62.361}
}

@article{metropolis,
  author  = {Hastings, W. K.},
  title   = {{Monte Carlo} sampling methods using {Markov} chains and their applications},
  journal = {Biometrika},
  volume  = {57},
  pages   = {97},
  year    = {1970},
  doi     = {10.1093/biomet/57.1.97}
}

@book{Schlkopf2001,
  author    = {Sch{\"o}lkopf, Bernhard and Smola, Alexander J.},
  title     = {Learning with Kernels: Support Vector Machines, Regularization, Optimization, and Beyond},
  publisher = {MIT Press},
  address   = {Cambridge, MA},
  year      = {2002}
}

@incollection{Rasmussen2004,
  author    = {Rasmussen, Carl Edward},
  title     = {{Gaussian} processes in machine learning},
  booktitle = {Advanced Lectures on Machine Learning},
  editor    = {Bousquet, Olivier and von Luxburg, Ulrike and R{\"a}tsch, Gunnar},
  series    = {Lecture Notes in Computer Science},
  volume    = {3176},
  publisher = {Springer},
  address   = {Berlin},
  pages     = {63},
  year      = {2004},
  doi       = {10.1007/978-3-540-28650-9_4}
}

@article{Mastromatteo_2011,
  author  = {Mastromatteo, Iacopo and Marsili, Matteo},
  title   = {On the criticality of inferred models},
  journal = {J. Stat. Mech.},
  volume  = {2011},
  pages   = {P10012},
  year    = {2011},
  doi     = {10.1088/1742-5468/2011/10/P10012}
}

@article{Janke_2004,
  author  = {Janke, W. and Johnston, D. A. and Kenna, R.},
  title   = {Information geometry and phase transitions},
  journal = {Physica A},
  volume  = {336},
  pages   = {181},
  year    = {2004},
  doi     = {10.1016/j.physa.2004.01.023}
}

@article{RevModPhys.67.605,
  author  = {Ruppeiner, George},
  title   = {{Riemannian} geometry in thermodynamic fluctuation theory},
  journal = {Rev. Mod. Phys.},
  volume  = {67},
  pages   = {605},
  year    = {1995},
  doi     = {10.1103/RevModPhys.67.605}
}

@article{Zdeborov__2016,
  author  = {Zdeborov{\'a}, Lenka and Krzakala, Florent},
  title   = {Statistical physics of inference: thresholds and algorithms},
  journal = {Adv. Phys.},
  volume  = {65},
  pages   = {453},
  year    = {2016},
  doi     = {10.1080/00018732.2016.1211393}
}

@article{Nguyen_Berg_2012,
  author  = {Nguyen, H. Chau and Berg, Johannes},
  title   = {{Bethe--Peierls} approximation and the inverse {Ising} problem},
  journal = {J. Stat. Mech.},
  volume  = {2012},
  pages   = {P03004},
  year    = {2012},
  doi     = {10.1088/1742-5468/2012/03/P03004}
}

\clearpage

\end{document}